\documentclass[11pt]{article}

\usepackage{amsthm}
\usepackage{amsmath}
\usepackage{amsfonts}
\usepackage{mathrsfs}
\usepackage{bm}
\usepackage{multirow} 
\usepackage{longtable}
\usepackage{geometry}
\usepackage{graphicx}
\usepackage{listings}
\usepackage{subfigure}
\usepackage{amssymb}
\usepackage{url}
\usepackage{xcolor} 
\usepackage[numbers, sort&compress]{natbib}
\usepackage{tabularx}
\usepackage{pdfpages}
\usepackage{emptypage}
\usepackage{subfigure}
\usepackage{float}
\usepackage{authblk}
\usepackage{makecell}

\newcommand{\RR}{\mathcal{R}}
\renewcommand{\AA}{\mathcal{A}}

\newcommand{\Bin}{\mbox{Bin}}
\newcommand{\Be}{\mbox{Be}}
\newcommand{\UPM}{\mbox{UPM}}
\newcommand{\RIntCRM}{\RR_{\mbox{Int-CRM}}}
\newcommand{\RmTPIt}{\RR_{\mbox{mTPI-2}}}
\newcommand{\RCCD}{\RR_{\mbox{CCD}}}

\title{A Unified  Decision  Framework for Phase I 
  Dose-Finding  Designs}

\author[1]{Yunshan Duan}
\author[2]{Shijie Yuan}
\author[3, *]{Yuan Ji}
\author[1]{Peter Mueller}
\affil[1]{{\small Department of Statistics and Data Science, University of Texas at Austin}}
\affil[2]{{\small Laiya Consulting, Inc., Shanghai, China}}
\affil[3]{{\small Department of Public Health Sciences, University of Chicago}}
\affil[*]{{\small Corresponding author:  Yuan Ji, YJi@health.bsd.uchicago.edu}}

\begin{document}
\maketitle

\newtheorem{theorem}{Theorem}
\newtheorem{prop}{Proposition}
\newtheorem{corollary}{Corollary} % [prop] % Use theorem counter as `parent`
\renewenvironment{proof}{\paragraph{Proof:}}{\hfill$\square$}
\newtheorem{lemma}{Lemma}
\newcommand{\dd}{\color{brown}}
\newcommand{\bb}{\color{black}}

\begin{abstract}
The purpose of a phase I dose-finding clinical trial is to investigate the
toxicity profiles of various doses for a new drug % dose toxicities
and identify the maximum tolerated dose. Over the past three decades,
various dose-finding designs have been proposed and discussed,
including conventional model-based designs, new model-based designs
using toxicity probability intervals, and rule-based designs. We
present a simple decision framework that can generate several popular
designs as special cases. We show that these designs share common elements under the framework, such as the same likelihood
function, the use of loss functions, and the nature of the optimal decisions as Bayes rules.
They differ mostly in the choice of the
prior distributions. We present theoretical results on the decision
framework and its link to specific and popular designs like mTPI,
BOIN, and CRM. These results provide useful insights into the
designs and their underlying assumptions, and convey information to
help practitioners select an appropriate design.
\end{abstract}

\maketitle

\section{Introduction}

We construct a Bayesian decision theoretic framework for dose finding
   trials and show how several popular designs can be derived as special
   cases. Understanding many designs as special cases of one common general
   construction helps investigators to put a rapidly increasing number of
   seemingly competing alternative designs into perspective and to
   appreciate the relative strengths and limitations of different
algorithms. A phase I clinical trial is the first stage of in-human investigation of
a new drug or therapy. Phase I dose-finding designs aim to
identify the maximum tolerate dose (MTD) and to provide dose
recommendation for later phase trials.  
In the vast majority of phase I trials, a set of ascending candidate doses is tested for toxicity and  the dose toxicity probability
is assumed to be monotonically increasing with the dose
level. Typically, the MTD is defined as the highest dose with a dose
limiting toxicity (DLT) probability closest to, or not higher than a
target toxicity probability $p_T$. Usually $p_T$ ranges from $0.17$ and $0.3$. In addition, some designs include the notion of an equivalence interval (EI)
to allow for 
variations in the definition of the MTD. For example, one may
choose to set $p_T = 0.3$ and EI $= (p_T - \epsilon_1, p_T +
\epsilon_2) = (0.25,0.35)$. This means that the target DLT probability
of the MTD is 0.3, but doses with DLT probabilities between 0.25 and
0.35 can also be considered as the MTD. In other words, the
EI allows investigators to consider doses with toxicity
probabilities within the EI interval as appropriate MTD
candidates.

\begin{figure}
	\centering
	\includegraphics[width=2.5in]{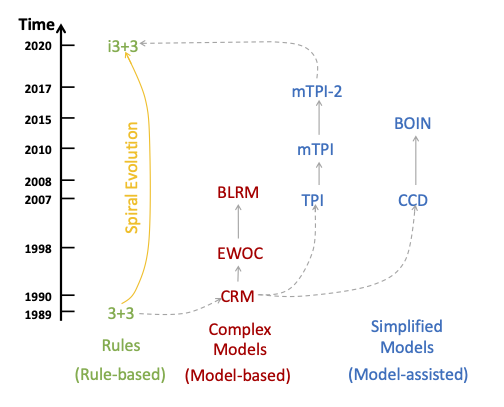}
	\caption{Illustration of some Phase I designs. Dotted lines connect designs across different categories, and solid lines connect designs within the same category. Arrows represents the sequence of original publication dates. }
	\label{fig:designs}
\end{figure}

A variety of statistical designs for phase I dose-finding trials has
been discussed in the literature. A design consecutively assigns
patients to recommended dose levels based on the observed DLT outcomes
from previously enrolled patients. Existing designs can broadly
be divided into two categories, rule-based designs and
model-based designs. Among model-based designs, some use simple models
and are sometimes called ``model-assisted'' designs. See Figure
\ref{fig:designs} for an illustration. We provide a brief introduction of the designs in Figure \ref{fig:designs} next.

The 3+3 design \citep{storer1989design} is
rule-based and consecutively assigns patients to the current dose or
the adjacent higher or lower doses based on observed DLT outcomes. For
example, if the current dose at which patients are assigned is $d$, then 3+3 assigns
the next patient cohort to doses $(d+1)$, $(d-1)$, or $d$
itself. This is called the ``up-and-down'' rule. Based on the same
up-and-down rule, a smarter rule-based design, i3+3, is proposed in
\citet{liu2020i3+} by accounting for higher sampling variability when the sample size at a dose is small. The i3+3 design maintains simplicity of rule-based designs, and exhibits operating characteristics comparable to more complex
model-based/assisted designs.

The continual reassessment method (CRM) \citep{o1990continual}, as the
first model-based design in the literature,
is based on an inference model with
a parsimoniously parameterized dose-response curve. During the trial,
CRM continuously updates the estimated dose-response curve based on the observed DLT data
throughout the trial. CRM has motivated subsequent important work on
model-based designs, including the Bayesian logistic regression method
(BLRM) in \citet{neuenschwander2008critical} and the escalation with
over-dose control (EWOC) in \citet{tighiouart2010dose}, among many
others, all leveraging parametric dose-response models for statistical
inference.

Recently, a class of designs, collectively known as ``interval-based
designs'' take advantage of the notion of an 
EI to simplify statistical modeling and
decision making for phase I trials. Notable examples include the
toxicity probability interval (TPI) design \citep{ji2007dose} and its
two modifications, mTPI \citep{ji2010modified}, mTPI-2
\citep{guo2017bayesian} (equivalently, Keyboard
\citep{yan2017keyboard}), the cumulative cohort (CCD) design
\citep{ivanova2007cumulative}, and the Bayesian optimal interval
(BOIN) design \citep{liu2015bayesian}. These designs use simple models
such as the beta/binomial hierarchical model and assume independence
across dose toxicity probabilities, without attempting to explicitly
model a dose-response curve.  While the independence model assumption is
apparently not true because dose toxicity is assumed to be
monotonically increasing, it does not affect the operating
characteristics of the interval-based designs due to various reasons
like the safety restrictions in practice (e.g., no skipping in dose
escalation). As a result, interval-based designs show robust
performance based on simple up-and-down rules, restricting dosing
decisions to be no more than one dose-level change from the current
dose used in the trial. In other words, a simple independent
beta/binomial model coupled with a simple up-and-down rule leads to
desirable simulation performance that justifies the application of
these designs. Importantly, the interval-based designs often generate
a decision table that greatly simplifies the trial conduct and allow
investigators to easily execute the dosing decisions provided in the
table.

In the past three decades, many designs (CRM, mTPI, mTPI-2, BOIN,
etc.) have been successfully applied to real-world trials. It is
natural to wonder which design or designs are suitable for a particular
trial. Recent reviews \cite{yuan2016bayesian} and \cite{horton2017performance}
provide some assessment of these designs, mainly from the perspective
of simulation performance. Occasionally, conflicting conclusions might
arise from different reviews based on the criteria used for design
evaluation, or from different scenarios considered in the
comparison. 
While simulation results can provide important information
on the numerical performance of the designs, we argue that a
theoretical investigation would complement the simulation results. In
this article, we show that using the same optimal decision rule under
the proposed decision framework, one can generate several published
designs as special cases. In other words, we show a theoretical
connection across different designs. These designs include mTPI,
mTPI-2, BOIN, and CCD.  In addition, based on the proposed decision
framework, we develop a new version of the CRM design, called Int-CRM,
that is founded on the same model assumption with the original CRM design but a different decision rule. We show that Int-CRM achieves comparable simulation performance as the original
CRM design and other interval-based designs. The general decision
framework provides insight into the similarities and
differences across various designs and may assist investigators to
select the right design for their specific needs.

The remainder of the paper is structured as follows. In Section
\ref{sec:framework}, we introduce the unified decision framework and its main components. Section \ref{sec:examples} shows how known designs fit into this framework, including the mTPI, mTPI-2, BOIN, CCD, and Int-CRM designs. In Section
\ref{sec:simulation}, we conduct simulation studies to assess the
operating characteristics of the designs using the i3+3 and CRM designs as benchmarks. Finally, we conclude
and end the paper with a discussion in Section \ref{sec:discussion}.

\section{ Decision Framework } \label{sec:framework}

\subsection{Overview} 
We cast the problem of dose finding as
an optimization in a decision problem. In particular we focus on the myopic decision problem of selecting the dose for the next
patient (cohort). 
It is myopic because the problem does not address the global decision of stopping the trial and dose selection; instead, the problem only considers the local optimal decision of finding the next dose for future patients.
The main components of the decision framework have been briefly illustrated in
\citet{guo2017bayesian} for the mTPI and mTPI-2 designs. 
A decision problem is characterized by an action space for decisions
$a$,
a probability model for all unknown quantities, and a loss function
\citep{berger2013statistical}.
Table \ref{tab:components} shows a summary of these components.

\begin{table}
	\centering
	 %\def\~{\hphantom{0}}
 %\begin{minipage}{175mm}
 \caption{Components of the proposed decision framework.}
 \label{tab:components}
	\begin{tabular}{ccc}
		 \hline \hline
		\textbf{Component}&\textbf{Notation}&\textbf{Notes}\\ \hline
		
          Probability model & $f(\bm{y} \mid \bm{\theta}) \pi(\bm{\theta} \mid m) \pi(m)$ 
                                 & {\small A hierarchical model with
                                   parameters $\bm{\theta}$ and $m$.}\\
                                    \hline
		Action &$a$& {\small Up-and-down dosing decisions, including}\\
                             & & {\small  $D$
                             (de-escalate), $S$ (stay), and $E$
                             (escalate).}\\
                              \hline
          Loss function&$\ell( a, \bm{\theta})$& {\small The loss for  taking
                                               action $a$ }\\
                               & & {\small where $\theta$ is the true parameter. } \\ \hline 
         Optimal rule&$\mathcal{R} = \arg \min_a \int \ell( a, \bm{\theta})p(\bm{\theta} \mid \bm{y})d\bm{\theta}$& {\small Bayes' rule that chooses the action } \\
         & & {\small with the minimal posterior expected loss. }\\ \hline \hline
	\end{tabular}
	%\end{minipage}
%\vspace*{2pt}
\end{table}

\subsection{General Framework}

In dose-finding trials with binary DLT endpoints, the parameter of
interest is   a set of   toxicity probabilities, $\bm{\theta} =
(p_1,...,p_T)$   at dose levels $x_d$, $d=1,\ldots,T$,
where $T$ is the number of dose levels,
and $p_d$ is
the toxicity probability at dose $d$. Let $y_d$ denote the number of
patients who experience DLTs out of $n_d$ patients treated at dose
$d$, and let $\bm{y} = (y_1,...,y_T)$.
  For all methods in the upcoming discussion  
the sampling model
$f(\bm{y} \mid \bm{\theta})$ in Table \ref{tab:components} is a
binomial distribution with parameter $p_d$, i.e.,
$$ y_d \mid p_d \sim \Bin(n_d,p_d), \;\;\; d=1,...,T,$$
implying a likelihood function,
$$ f(\bm{y} \mid \bm{\theta}) \propto \prod_{d=1}^T p_d^{y_d} (1-p_d)^{n_d-y_d} .$$
For model-based designs with a dose-response curve, toxicity
probabilities are modeled as a function of the dose levels $x_d$. For
example, a version of the CRM assumes $p_d = q_{d}^{exp(\alpha)}$, and
a single parameter $\theta= \alpha$ (the values $q_{d}$ are fixed
  and are known as the ``skeleton'').
The BLRM design uses $p_d = logit^{-1}(\alpha + \beta x_d)$ with
parameters $\bm{\theta} = (\alpha, \beta)$.

The proposed decision framework uses a concept of probability
intervals. The parameter of interest is $p_d$, and the parameter
space of $p_d$ is $I= [0,1]$. Consider a set of intervals within $I$,
denoted as $\Omega = \{I_k,\; k=1,...,K\}$,   which form   a partition of the
parameter space $I$. That is, $\bigcup_{k=1}^K I_k = I$ and $I_k \cap
I_{k'} = \emptyset, \;\; k \neq k'$. The true value of $p_d$ belongs
to one and only one of the intervals. For example, $\Omega = \{I_1 =
[0,0.5], I_2=(0.5,1]\}$ is a partition, and if $p_d= 0.3$, $p_d \in
I_1$. We introduce a latent indicator $m_d$ (or, for short, just $m$) with $m_d=k$ if $p_d \in I_k$,
and define a hierarchical model prior $\pi(m)$ and $\pi(p_d \mid m)$. For
example, $\pi(m=k) = \frac{1}{K}, \;\;\; k=1,...,K,$ and  
$\pi(\bm{\theta} \mid m=k) \propto \prod_{d=1}^{T} \Be(\alpha,\beta)
\delta(p_d \in I_k),$ a truncated beta distribution.
Here, $\delta(\cdot)$ is an indicator function.
That is, $p_d$ are conditionally independent with pdf  
$$
  p(p_d \mid m=k) = \frac{beta(p_d;\alpha,\beta)\delta(p_d \in I_k)}{\int_{I_k}
  beta(p_d;\alpha,\beta) dp_d},
$$
where $beta(p_d;\alpha,\beta) =
\frac{\Gamma(\alpha+\beta)}{\Gamma(\alpha)\Gamma(\beta)}p_d^{\alpha-1}(1-p_d)^{\beta-1},
\alpha>0, \beta>0$   is a $\Be(\alpha,\beta)$ p.d.f.  

We consider a special partition $\Omega = \{I_1, I_2, I_3\}$ where
$I_2 = I_S = EI = (p_T - \epsilon_1, p_T + \epsilon_2)$, $I_1
= I_E = [0,p_T - \epsilon_1]$, and $I_3 = I_D = [p_T
+ \epsilon_2,1]$. Therefore, $K=3$ and we use notations $I_S$, $I_E$,
and $I_D$ to associate the intervals with corresponding up-and-down
dose-finding decisions S, E, and D, respectively.
We summarize the proposed decision framework below.

\smallskip

\paragraph{Likelihood}

\begin{equation}
    f(\bm{y} \mid \bm{\theta}) \propto \prod_{d=1}^{D} p_d^{y_d} (1-p_d)^{n_d-y_d},
    \label{likelihood}
\end{equation}
where $p_d$ is the toxicity probability for dose $d$, $d=1, \dots, D$.

\paragraph{Prior}
We assume $p_d$ are {\em a priori} independent and 
\begin{eqnarray*}
\pi(p_d \mid m=k)  &\propto& g(p_d) \delta(p_d \in I_k), \;\;\; k=1,...,K, \\
\pi(m = k)  &=& \frac{1}{K}, \;\;\; k=1,...,K.
\end{eqnarray*}
For example, $g(p_d) = beta(p_d;\alpha,\beta)$.

\smallskip

\paragraph{Partition}

$\Omega = \{I_E (I_1), \; I_S (I_2), \; I_D (I_3)\}$,
where $I_E = [0,p_T-\epsilon_1], I_S = (p_T-\epsilon_1,p_T+\epsilon_2)$, and $I_D = [p_T+\epsilon_2,1]$, and $I_1 = I_E$, $I_2=I_S$, and $I_3=I_D$.

\paragraph{Actions} The actions are the three up-and-down decisions
for dose-finding, i.e., 
$$a \in \AA = \{E,S,D\},$$
where $\AA$ denotes the action space. Here $E$, $S$, $D$
denote the dosing decisions ``Escalation'', ``Stay'', and
``De-escalation,'' respectively. In particular, if the last patient was
assigned dose $d$, then $E$, $S$, or $D$ means treating future
patients at dose $(d+1)$, $d$, or $(d-1)$, respectively. 

\smallskip

\paragraph{Loss}
  We proceed with a myopic perspective, focusing on the decision
for the respective next patient (cohort), and therefore specify a loss
function for the next dose assignment $a$ only.

We use a 0-1 loss function,
\begin{eqnarray} \label{eq:loss}
\ell(a,p_d) = 
 \left\{\begin{array}{ll}
1, & p_d \notin I_a \\
0, & p_d \in I_a
\end{array}\right. , \;\;\;\;  a \in \AA = \{E,S,D\}.
\end{eqnarray}
In words, when the action corresponds to an interval which contains the true parameter, the loss takes the value 0; otherwise, the loss equals 1. The loss function $\ell(a,p_d)$ is stated in Table \ref{tab:mptiloss}. 

\begin{table}[b]
 \vspace*{-6pt}
 \centering
 \def\~{\hphantom{0}}
	\caption{The 0-1 loss function $\ell(a,\theta)$.}
	\label{tab:mptiloss}
	\begin{tabular}{c|ccc}
		\hline \hline
		& \multicolumn{3}{c}{$p_d \in$} \\
		$\ell(a,p_d)$ & $[0,p_T-\epsilon_1]$ & $(p_T-\epsilon_1,p_T+\epsilon_2)$ & $[p_T+\epsilon_2,1]$ \\ \hline
		a = $D$ &1 & 1 &0 \\
		a = $S$ & 1 & 0 & 1\\
		a = $E$ & 0 & 1 & 1\\
		\hline \hline
	\end{tabular}
	\vskip18pt
\end{table}

In other words, the loss function $\ell(a,p_d)$ defines a 0-1 estimation loss for $m$,
i.e., the interval that contains $p_d$.

Two more comments about the loss function and the setup of the
  decision problem.
  First, in general a loss (or, equivalently, utility) function could
  also be an argument of the outcome $y_d$. This is relevant, for
  example, if instead of inference loss we focus on the patients'
  preferences. However, the intention of this discussion is only to
  highlight common structure in existing dose finding methods, for
  which we only need this restricted inference loss.
  Another important limitation is the myopic nature of the setup.
  We consider the dose allocation for each patient (or patient cohort)
  in isolation, ignoring that dose allocation now might help later
  decisions. That is, we ignore the sequential nature of the problem.
  Again, for the upcoming exposition of common underlying structure for
  the considered dose finding methods we will only refer to this myopic
  decision problem.
  
\smallskip

\paragraph{Bayes' rule} 
The optimal decision rule for dose $d$ is the Bayes' rule, defined as
\begin{eqnarray} \label{eq:bayesrule}
\mathcal{R}_d = \arg \min_{a \in \AA} \int \ell(a,p_d) p(p_d \mid \bm{y}) dp_d, 
\end{eqnarray}
which minimizes the posterior expected loss. Here, $p(p_d \mid \bm{y})$ is the posterior distribution of $p_d$.

\medskip
In general, under a 0-1 estimation loss for a discrete parameter the Bayes rule is simply the posterior mode. The following result states this in the context of our problem. The Bayes' rule is equivalent to the result of finding the interval with the maximal posterior probability.

\medskip

% \paragraph{Result 1.}
\paragraph{Proposition 1}
%\begin{prop} \label{result1}
Denote $\Omega = \{I_1,I_2,I_3\} = \{I_E,I_S,I_D\}$, where
$I_1 = I_E$, $I_2 = I_S$, $I_3 = I_D$. Suppose dose $d$ is the current
dose. Let $\{m=k\}$ denote the event $\{p_d \in I_k\}, \; k
\in \{1,2,3\}$. Let $\AA = \{E,S,D\}$. Assume $\pi(m=k) =
\frac{1}{3}, \; k \in \{1,2,3\}$. The Bayes' rule under the 0-1 loss
in Equation (\ref{eq:loss}) is given by
\begin{eqnarray} \label{eq:the1}
	\mathcal{R}_d & = & \arg \max_{a \in \AA} Pr(p_d \in I_a
                        \mid \bm{y}) \\ \nonumber
	 &= & \arg \max_{k \in \{1,2,3\}} Pr(m=k \mid \bm{y}) 	
\end{eqnarray}
%\end{prop}
\noindent   See Appendix C for a proof.

\section{Design Examples} \label{sec:examples}
  We show how various designs fit as special cases into this
framework. That is,  
we provide examples of the decision framework that give rise to
well-known designs including mTPI, BOIN, CCD, mTPI-2, and a new
version of CRM, called the Int-CRM design.

\subsection{Interval-based designs}
We first introduce the connection between the decision framework and
the interval-based designs, mTPI, mTPI-2, BOIN and
CCD. These designs share some common components under the framework, but
also   include some elements specific to each design.  

\paragraph{Common components} Likelihood (\ref{likelihood}), prior $\pi(m)$, loss
function (\ref{eq:loss}), and the nature of the defined dose allocation as Bayes' rule. 

\paragraph{Individual components} Prior $\pi(p_d \mid m)$, the
specific partition
$\Omega = \{I_k,\; k=1,...,K\},$ and the definition of the action set $\AA$.

\medskip
All four interval-based designs use the binomial sampling model (\ref{likelihood}).
And the designs share the same discrete uniform
prior $\pi(m)$, the 0-1 loss function and the   use of Bayes' rule
to select a decision.  
They divide the $[0,1]$ parameter space of $p_d$
into different intervals and use different priors. See Table
\ref{tab:ind_components} as a summary. We discuss details for each design
next.

\newcommand{\tabincell}[2]{\begin{tabular}{@{}#1@{}}#2\end{tabular}}

\begin{table}
	 \centering
 %\def\~{\hphantom{0}}
% \begin{minipage}{130mm}
 \caption{Individual components of the proposed decision framework for some interval-based designs.}
	\label{tab:ind_components}
	%\begin{center}
	\begin{tabular}{c|c|c|c}
		\hline \hline
		 &\textbf{mTPI}&\textbf{mTPI-2}&\textbf{BOIN/CCD}\\ \hline
		 
		 Actions & $\AA = \{E,S,D\}$ & $\AA = \{1,...,K\}$&$\AA = \{E,S,D\}$ \\ \hline
		 
		  & $I_E = [0,p_T-\epsilon_1]$ & $I_E = I_{E,1} \cup \cdots \cup I_{E,K_1} $ & $I_E = [0,\phi_E]$ \\
		  Intervals & $I_S = (p_T-\epsilon_1,p_T+\epsilon_2)$ & $I_S = (p_T-\epsilon_1,p_T+\epsilon_2)$ & $I_S = (\phi_E,\phi_D)$ \\
		  & $I_D = [p_T+\epsilon_2,1]$ & $I_D = I_{D,1} \cup \cdots \cup I_{D,K_2}$ & $I_D = [\phi_D,1]$ \\ \hline
		  Priors & $\pi(p_d \mid m=k) \propto $ & $\pi(p_d \mid m=k) \propto $ & $\pi(p_d \mid m=k) = $ \\
		  & $Be(1,1) \delta(p_d \in I_k)$ & $Be(1,1)\delta(p_d \in I_k)$ & {$\delta(p_d = \phi_k)$}\footnote{See Theorem \ref{theorem2} for details. } 
		  \\ \hline
		  \footnotesize{$^1$See Theorem \ref{theorem2} for details. }
	\end{tabular}
	%\end{center}
%	\end{minipage}
%	\vskip18pt
\end{table}

\bigskip

\underline{The mTPI design}

For the mTPI design, given the equivalence interval
EI$=(p_T-\epsilon_1,p_T+\epsilon_2)$, the $[0,1]$ parameter space is
naturally partitioned into three intervals $\Omega = \{I_1 = I_E =
[0,p_T-\epsilon_1], I_2 = I_S = (p_T-\epsilon_1,p_T+\epsilon_2), I_3 =
I_D = [p_T+\epsilon_2,1]\} $ that correspond to the actions $\AA =
\{E,S,D\}$, as shown in Table \ref{tab:ind_components}. The mTPI
decision is equivalent to the Bayes' rule $\RR_d$ under the decision
framework. See Corollary \ref{cor:mtpi} below for a formal mathematical description.

\bigskip

\begin{corollary}
  \label{cor:mtpi}
The mTPI decision in \citet{ji2010modified} is given by $$R_{mTPI} =
\arg \max_{a \in \{E,S,D\}} \UPM(I_a),$$ where UPM stands for ``unit
probability mass'' and $\UPM(I_a) = Pr^*(p_d \in I_a) / ||I_a||$; here
$||I_a||$ is the length of $I_a$, and 
$Pr^*(p_d \in I_a) = \int
B(y_d+1,n_d-y_d+1) \cdot p_d^{y_d}(1-p_d)^{n_d-y_d} \cdot \delta(p_d
\in I_a) dp_d$
is calculated based on $p_d \sim \Be(y_d+1,
n_d-y_d+1)$. Let $I_1 = I_E = [0,p_T-\epsilon_1]$, $I_2 = I_S =
(p_T-\epsilon_1,p_T+\epsilon_2)$, and $I_3 = I_D =
[p_T+\epsilon_2,1]$. Then $R_{mTPI} = \RR_d$, the Bayes rule under
$$\pi(p_d \mid m = k) = C_k \cdot beta(p_d;1,1)\delta(p_d \in
I_k),$$
where $beta(\cdot;1,1)$ denotes the density function of $\Be(1,1)$
distribution and $ C_k = \frac{1}{\int_{I_k} beta(p;1,1) dp}$
is a normalizing constant.  
\end{corollary}
\noindent   See Appendix C for a proof.  

\bigskip

\underline{The mTPI-2 design}

Ockham's razor is a principle   in statistical inference calling
for  
an explanation of the facts to be no more complicated than necessary
\citep{thorburn1918myth, jefferys1992ockham}. In
  the context of    model
selection, the Ockham's razor prefers parsimonious models that
describe the data equally well as more complex models. In the proposed
decision framework, $\{m=k\}, \; k=1,...,K$, is equivalent to $K$
models $\{M_k: m = k\}$, and choosing the value of $m$ is equivalent
to a model selection problem. Bayesian model selection chooses the
model with the largest posterior probability
(compare Proposition 1), i.e., $Pr(m=k \mid y)$,
and models are automatically penalized for their complexity. In other
words, Bayes' rule $\RR_d = \arg \max_{k \in \{1,2,3\}} Pr(m=k
\mid y)$ implements Ockham's razor   if we define   model complexity as
$||I_k||$. Therefore, when the three models are $I_1= I_E =
[0,p_T-\epsilon_1]$, $I_2 = I_S = (p_T-\epsilon_1, p_T+ \epsilon_2)$,
$I_3 = I_D = [p_T+\epsilon_2]$, the ``simplest'' model is $I_2=I_S$,
since $\epsilon_1$ and $\epsilon_2$ are typically small probabilities
($\leq 0.05$).

\citet{guo2017bayesian}   explain   mTPI-2 as aiming to blunt
Ockham's razor by redefining a (finer) partition $\Omega^* = \{I_E^*,I_S = EI, I_D^*\},$ where
$I_E^* = \{I_{E,1}, ... , I_{E,K_1}\} $ and $I_D^* = \{I_{D,1}, ...,
I_{D,K_2}\}$. Probability intervals $\{I_{E,k}\}_{k=2}^{K_1}$ and
$\{I_{D,k}\}_{k=2}^{K_2}$ have the same length as $I_S = EI$. Let $K =
K_1 + K_2 +1.$
The   selected   model $m$ under the mTPI-2 design can then be
shown to be Bayes' rule $\RR_d$,   under an action set   $\AA_m =
\{1,...,K\}$. Corollary \ref{cor:mtpi2} next summarizes the results.

% %Rather than assuming a same loss for $p_d$ mistakenly falling into
% each intervals like the mTPI design, mTPI-2 divides the $(0,1)$
% toxicity into several intervals with same lengths as the equivalence
% interval $(p_T-\epsilon_1,p_T+\epsilon_2)$. Denote the $K$ intervals
% as$$ I_k, \;\; k=1,...,K.$$ The optimal action $a$ of mTPI-2 design is
% the Bayes' rule $R$ and it takes value from $\AA =
% \{1,...,K\}$.Corollary \ref{cor:mtpi2} shows that when the
% corresponding up-and-down decision is

\begin{corollary}
	\label{cor:mtpi2}
	Under mTPI-2, $\Omega^* = \{I_E^* = \{I_{E,1}, \cdots,
        I_{E,K_1}\}, I_S = (p_T-\epsilon_1,p_T+\epsilon_2), I_D^* = \{
        I_{D,1}, \cdots ,I_{D,K_2}\} \}$, $\AA_m =
        \{1,...,K\}$. Assume the prior on $p_d$ is conditionally independent and
        given by $$\pi(p_d \mid m = k) = C_k \cdot
        beta(p_d;1,1)\delta(p_d \in I_k).$$ 
	Then Bayes' rule is
        $$
	\RR_d  = \arg \max_{m \in \AA_m} Pr(m=k \mid y) 
	 = \arg \max_{m \in \AA_m} \UPM(I_m).
	$$
\end{corollary}
\noindent   See Appendix C for a proof.  
Corollary \ref{cor:mtpi2} establishes the Bayes' rule $\RR_d$ as
an action in $\AA_m = \{1,...,K\}$. To see the
connection to the dose-finding decision in mTPI-2, we refer to the
next result.
% simple Corollary \ref{cor:mtpi2_2} below. 

\bigskip

\begin{corollary}
	\label{cor:mtpi2_2}
	Let 
	\begin{eqnarray*} 
	\RmTPIt =
	 \left\{\begin{array}{ll}
	E, &  \arg \max_{I_m} \UPM(I_m) \subset (0,p_T-\epsilon_1), \\
	S, &  \arg \max_{I_m} \UPM(I_m) = (p_T-\epsilon_1,p_T+\epsilon_2), \\
	D, &  \arg \max_{I_m} \UPM(I_m) \subset (p_T+\epsilon_2,1).
	\end{array}\right.
	\end{eqnarray*}
	  Then   
	\begin{eqnarray*} 
          { \RmTPIt} =
	 \left\{\begin{array}{ll}
	E, &  I_{\RR_d} \subset (0,p_T-\epsilon_1), \\
	S, &  I_{\RR_d} = (p_T-\epsilon_1,p_T+\epsilon_2), \\
	D, &  I_{\RR_d} \subset (p_T+\epsilon_2,1).
	\end{array}\right.
	\end{eqnarray*}
\end{corollary}

\bigskip

In other words, if $\RR_d = m^*$, then $I_{m^*}$ is the interval with the largest UPM, for $m^* \in \AA_m$. And if $I_{M^*}$ is below, equal to, or above the EI$=(p_T-\epsilon_1,p_T+\epsilon_2)$, the decision is $E$, $S$, or $D$, respectively. This is the same as the up-and-down rule in the mTPI-2 design in \citet{guo2017bayesian}. Proof of Corollary \ref{cor:mtpi2_2} is immediate and omitted.

\bigskip

\underline{The BOIN design}

The BOIN design \citep{liu2015bayesian} uses a decision rule 
\begin{eqnarray*}
R_{BOIN} =
 \left\{\begin{array}{ll}
E, & \hat{p}_d \leq \lambda_1, \\
S, & \lambda_1 < \hat{p}_d < \lambda_2, \\
D, & \hat{p}_d \geq \lambda_2 ,
\end{array}\right.
\end{eqnarray*}
where $ \hat{p}_d = \frac{y_d}{n_d} $,
$
\lambda_1 = \xi(\phi_E;\phi_S), 
\lambda_2 = \xi(\phi_D;\phi_S),
$
and
\begin{equation}
  \xi(\phi_i;\phi_j) = \frac{log\left(
      \frac{1-\phi_i}{1-\phi_j}\right) }{log\left(
      \frac{\phi_j(1-\phi_i)}{\phi_i(1-\phi_j)}\right) }.
  \label{eq:xi}
\end{equation}  
In particular, $\phi_S = p_T$, $\phi_E (< p_T)$ and $\phi_D (>p_T)$
are pre-specified values. Here, $\phi_E$ and $\phi_D$ play a similar
rule as $(p_T - \epsilon_1)$ and $(p_T + \epsilon_2)$ in the mTPI and
mTPI-2 designs, which defines the boundaries of an initial equivalence interval elicited from clinicians. We show that
the decision rule $R_{BOIN}$ is also a Bayes' rule under the
proposed decision framework next. 

\bigskip

\begin{theorem} \label{theorem2}
	Assume $y_d \mid p_d \sim \Bin(n_d,p_d)$, $\pi(p_d  \mid m=k)
        = \delta(p_d = \phi_k)$, for $k=1,2,3$, and $\phi_1 = \phi_E$,
        $\phi_2 = \phi_S = p_T$, and $\phi_3 = \phi_D$. Under the 0-1
        loss $\ell(a, p_d)$ in Equation (\ref{eq:loss}), the Bayes' rule is
        equivalent to $R_{BOIN}$, i.e., 
	\begin{eqnarray*}
	\RR_d = R_{BOIN} =
	 \left\{\begin{array}{ll}
	E, & \hat{p}_d \leq \lambda_1, \\
	S, & \lambda_1 < \hat{p}_d < \lambda_2, \\
	D, & \hat{p}_d \geq \lambda_2 ,
	\end{array}\right.
	\end{eqnarray*}
	where $ \hat{p}_d = \frac{y_d}{n_d} $,
	$
	\lambda_1 = \xi(\phi_E;\phi_S), 
	\lambda_2 = \xi(\phi_D;\phi_S),
	$
	and $\xi$ is defined as in Equation (\ref{eq:xi})
	% $$ \xi(\phi_i;\phi_j) = \frac{log\left( \frac{1-\phi_i}{1-\phi_j}\right) }{log\left( \frac{\phi_j(1-\phi_i)}{\phi_i(1-\phi_j)}\right) }. $$
\end{theorem}
  See Appendix C for a proof.  
By Theorem \ref{theorem2} the BOIN design   takes the form of the  
Bayes' rule under the same decision framework using the 0-1 loss. BOIN
uses a point-mass prior for $p_d$ on three values, $\phi_E$, $\phi_S$,
$\phi_D$, while mTPI/mTPI-2 using truncated beta priors instead. Next,
we show that the BOIN design is almost the same as the CCD design.
  This is easiest seen   
under the perspective of the proposed decision framework. The
difference between the two designs are the locations of the point-mass
priors. 

\bigskip

\underline{The CCD design}

The CCD design \citep{ivanova2007cumulative} compares $\hat{p}_d$ with $(p_T - \epsilon_1)$ and $(p_T + \epsilon_2)$, and uses the following up-and-down rule,
\begin{eqnarray*}
\RCCD &=&
 \left\{\begin{array}{ll}
E & \hat{p}_d \leq p_T - \epsilon_1 \\
S & p_T - \epsilon_1 < \hat{p}_d < p_T + \epsilon_2 \\
D & \hat{p}_d \geq p_T + \epsilon_2
\end{array}\right.
\end{eqnarray*}
Corollary \ref{cor:ccd} shows that the decision of the CCD design is the same Bayes' rule in the same framework as BOIN but with a different prior distribution.

\begin{corollary}
  \label{cor:ccd}
  The CCD decision $\RCCD = \RR_d$ with $$\pi(p_d \mid m = k) = \delta(p_d = \phi'_k), \;\;\; k= E,S,D,$$ where 
  $\phi'_E = \xi^{-1} (p_T-\epsilon_1)$, 
  $\phi'_D = \xi^{-1} (p_T+\epsilon_2)$,
  $\phi'_S = \phi_S = p_T$,
  and $\xi(\phi) \equiv \xi(\phi, p_T)$ in Equation (\ref{eq:xi}).
  % $$ \xi(\phi) = \frac{log\left( \frac{1-\phi}{1-p_T}\right) }{log\left( \frac{p_T(1-\phi)}{\phi(1-p_T)}\right) }. $$
\end{corollary}
\noindent   See Appendix C for a proof.  

Corollary \ref{cor:ccd} shows that BOIN and CCD are identical designs
with the only difference being that BOIN uses a point-mass prior
$\pi(p_d \mid m=k) = \delta(p_d=\phi_k)$, whereas CCD uses $\pi(p_d
\mid m=k)= \delta(p_d=\phi'_k)$.

\bigskip

\underline{The Int-CRM design}

Using the same decision framework, we propose a variation of the CRM
design, called Int-CRM.
We assume the same parametric dose-response model as in the CRM design
\citep{o1990continual},   with   the probability of toxicity
monotonically increasing with dose. Let $d_{i}$ denote the dose 
for the $i$th patient, $d_i \in \{1,...,T\}$,
and $Y_i$ the
binary indicator of DLT. The dose-response curve is assumed to be the
power model as in the CRM, 
$$ F(d,\theta) = q_d^{exp(\theta)}, $$
where $(q_1,..,q_T)$ (``skeleton'') are {\em a priori} pre-specified dose
toxicity probabilities. Other sensible dose-response models, such as a logit model, may be considered as well.  The toxicity rates
are dependent across doses through the dose-response curve and the
inference is based on the parameter $\theta$. The likelihood function is given by
$$
f(\bm{y} \mid \theta) \propto  \prod_{i=1}^n F(d_{i},\theta)^{y_i}
\{1- F(d_{i},\theta)\}^{1-y_i},
$$ 
where $n$ is the number of patients in the trial.

Following \citet{cheung2002simple}, we define an interval
$[A_1,A_{ T+1}]$   for $\theta$ that is   wide enough to
allow for a wide range of dose-response curves. For example, set
$A_1$ and $A_{ T+1}$ so that $q_1^{exp(A_1)} > 1-10^{-5}$ and
$q_ T^{exp(A_{ T+1})} < 10^{-5}$, which correspond to response curves
constantly equal to 1.0 and 0.0, respectively, and $\theta \in
[A_1,A_{ T+1}]$ allows choices in-between these extremes. Using $A_1$
and $A_{ T+1}$, we define sub-intervals for $\theta$ as the set of
values that imply $d_k$   having toxicity probability   closest to $p_T$,
$$ I_k = \{\theta \in [A_1,A_{ T+1}]: |F(k,\theta) - p_T| < |F(d,\theta) - p_T|, \forall d \neq k \},$$ k=1, \dots , T. 
  As shown in   \citet{cheung2002simple}, $I_k$ is an interval,
denoted as $I_1 = [\psi_1 = A_1, \psi_2)$, $I_k = [\psi_k,\psi_{k+1}),
\; k=2,...,( T-1)$, $I_{ T} = [\psi_ T,\psi_{ T+1} = A_{ T+1}]$, where
$\psi_k$ is implicitly defined as the solution of
$$ F(k-1,\psi_k) + F(k, \psi_k) = 2p_T, \;\;\; k=2,..., T.$$
Given the ``skeleton'' $(q_1,...,q_T)$, we can obtain the numerical result of the interval boundaries $\phi_k$'s by solving the equation above. See Appendix B for details. Each interval consists of a set of $\theta$ values where dose $k$'s
toxicity probability is the closest to $p_T$ among all the doses. We
use these intervals $I_k$'s in our framework for Int-CRM. We propose
hierarchical priors

$$\pi(m=k) = \frac{1}{ T}, \;\;\; k=1,...,T$$ and $$\pi(\theta \mid m = k) = \frac{\phi(\theta) \delta(\theta \in I_k)}{\int_{I_k}\phi(\theta) d\theta},$$
where $\phi(\theta)$ is the density function of the normal
distribution $N(0,\sigma^2)$.

The action space of the Int-CRM design is $\AA = \{1,..., T\}$,
corresponding to the dose for treating the next patient. Following the
proposed decision framework, we use the 0-1 loss function and the
Bayes' rule that minimizes the posterior expected loss for the Int-CRM
decision.

\begin{theorem}
	\label{theoremInt-CRM}
	Under the 0-1 loss, i.e.,
	\begin{eqnarray*} 
	\ell(a,\theta) = 
	 \left\{\begin{array}{ll}
	1, & \theta \notin I_a \\
	0, & \theta \in I_a
	\end{array}\right., \;\;\;\;  a \in \AA = \{1,..., T\},
	\end{eqnarray*}
	the Int-CRM decision is the Bayes' rule
	\begin{eqnarray*}
	\RIntCRM &=& \arg \max_{k \in \AA } Pr(m=k \mid y) \\
	  &=& \arg \max_{k \in \AA} \int  p(y \mid \theta)\pi(\theta \mid m=k) d\theta \\
	&=& \arg \max_{k \in \AA} \int \prod_{i=1}^n F(d_{i},\theta)^{y_i} \{1- F(d_{i}, \theta)^{1-y_i}\} \frac{\phi(\theta) \delta(\theta \in I_k)}{\int \phi(\theta)\delta(\theta \in I_k) d\theta} d\theta.
	\end{eqnarray*}
\end{theorem}
The proof is immediate by the definition of Bayes' rule. Below is the proposed Int-CRM dose-finding algorithm.

\medskip

%\clearpage \newpage
{
  \noindent \fbox{%
  \centering
    \parbox{0.9\textwidth}{
    {\small
      \textbf{{\normalsize The Int-CRM Algorithm:} }
      
      \vskip 0.1in 
      \textbf{Dose Finding Rules: } 
      After each cohort of patients completes the DLT follow-up
      period,  the dose to be assigned is the $\RIntCRM$, the Bayes'
      rule, unless the following safety rules apply. 
      
      \vskip 0.1in 
      \noindent \textbf{Safety Rules: }  Four additional
      rules are applied for safety.  
      \begin{itemize}
      \item[] \underline{Rule 1: Dose Exclusion:} If the current dose is
        considered excessively toxic, i.e., $Prob\{ p_d>p_T\mid data
        \}>\xi$ (see below about evaluating this probability),
        where the threshold $\xi$ is close to 1, say 0.95,
        the current and all higher doses will be excluded in the
        remainder of the trial to avoid assigning any patients to
        those doses.  
      \item[] \underline{Rule 2: Early Stopping:} If the current dose is the
        lowest dose (first dose) and is considered excessively toxic,
        i.e., $p\{ p_1>p_T\mid data \}>\xi$, where the threshold
        $\xi$ is close to 1, say 0.95, stop the trial early and
        declare no MTD. 

          To evaluate $p\{ p_d>p_T\mid data \}$ in Rules 1 and 2 we
        use a $\Be(\alpha_0+y_d,\beta_0+n_d-y_d)$ distribution with
        $\alpha_0=\beta_0=1$.  
        
      \item[] \underline{Rule 3: No-Skipping Escalation:}
        If the dose-finding rule recommends escalation, such escalation shall not increase the dose by more than one level. Dose-escalation cannot increase by more than one level. 
        That is, suppose the current dose
        is $d$. 
        If the next recommended dose $\RIntCRM$   is such that  
        $(\RIntCRM-d)>1$,
        escalate to dose $(d+1)$ instead.  
        
      \item[] \underline{Rule 4: Coherence:}
        No escalation is permitted if the empirical rate of DLT for
        the most recent cohort is higher than $p_T$, according to the
        coherence principle \citep{cheung2011dose}.  
      \end{itemize}
      
      \vskip 0.1in 
      \noindent \textbf{Trial Termination: } The trial
      proceeds unless any of the following stopping criteria is met:  
      \begin{itemize}
      \item If the pre-specified maximum total sample size $n$ is reached. 
      \item Rule 2 above.
        
      \end{itemize} 
      
      \vskip 0.1in 
      \noindent \textbf{MTD Selection: } Once all the
      enrolled patients complete the DLT observation and the trial is
      not stopped early, the last dose level $\RIntCRM$ is selected as
      the MTD. 
      }
      
    }
  }
}

\clearpage \newpage

\section{Simulation Studies} \label{sec:simulation}

\subsection{Simulation Settings}

  We set up simulation studies to evaluate   the operating
characteristics of
  the different designs that we have shown to be special cases of
  the proposed general framework, 
including the mTPI, mTPI-2, BOIN, CCD and the Int-CRM designs. We show how the common underlying decision framework leads to very similar performances of the designs under consideration. We also compare to the i3+3 design and
the original CRM design as benchmarks. 

\smallskip

\paragraph{Fixed Scenarios} We use a total of 15 scenarios, with a
set of  $ T= 4,5,$ or $6$ doses.
Assume the target toxicity probability
$p_T = \phi_S = 0.3$ ($\phi_S$ is the notation in BOIN), and   maximum   sample size of 30.  
For all designs we apply the same safety rules as in the mTPI, mTPI-2,
and Int-CRM designs. See Appendix  A for details.  
For interval-based designs we use EI $= (p_T -\epsilon_1,p_T +
\epsilon_2)$, $\epsilon_1 = \epsilon_2 = 0.05$. 
For the Int-CRM and CRM design, the skeleton $q_d$ is generated using
the approach proposed in \citet{lee2011calibration}, which selects the
skeleton based on indifference intervals for the MTD. Also, we set the
half width of the indifference intervals, $\delta = 0.05$. We apply the
coherence principle \citep{cheung2011dose}, avoiding
immediate escalation after toxic outcomes. 

For the BOIN design, we set $\lambda_1 = p_T -\epsilon_1$,  $\lambda_2
= p_T +\epsilon_2 $. This is equivalent to setting  $\phi_E = \xi^{-1}
(p_T-\epsilon_1)$, and $\phi_D = \xi^{-1} (p_T+\epsilon_2)$. 
By Theorem \ref{theorem2}, these   values for   $\lambda_1$ and
$\lambda_2$  
make the BOIN decision identical to the CCD design, leading to same
operating characteristics of the two designs.

\smallskip

\begin{figure}[h]
	\centering
	\includegraphics[width=3in]{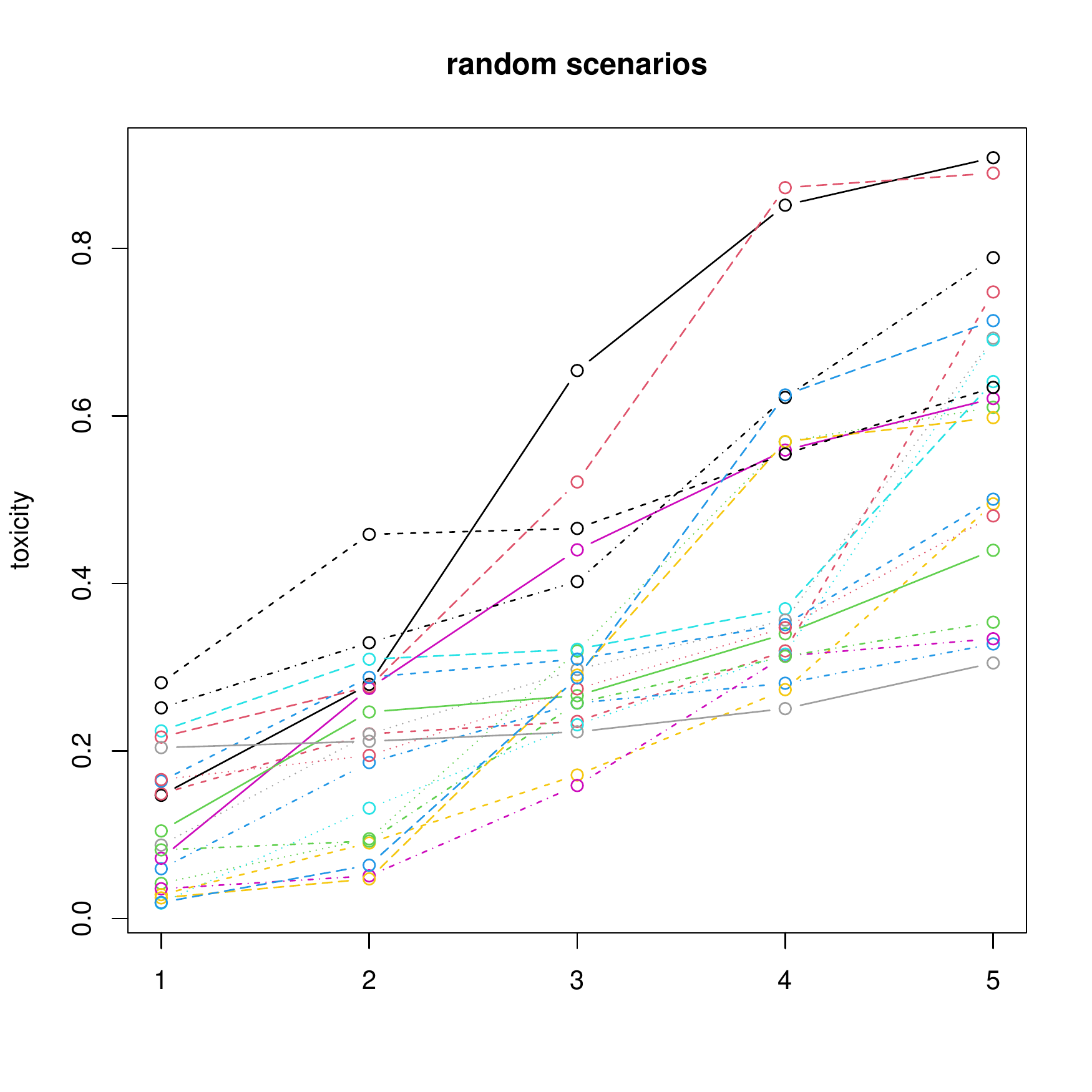}
	\caption{  Illustration of the first 20 random scenarios of true toxicity probabilities in the simulation.  }
	\label{fig:ran_scenarios}
\end{figure}

\paragraph{Random Scenarios} We generate additional 1,000 random scenarios
to further evaluate the designs. Scenarios are generated based on the
pseudo-uniform algorithm in \citet{clertant2017semiparametric}. Figure
\ref{fig:ran_scenarios} plots the first 20 scenarios.
Other settings of the designs are the same as the fixed
scenarios, such as $p_T$ and EI, $\lambda_1, \lambda_2$ for the BOIN
design, and $\delta$ for the Int-CRM and CRM designs. 

\subsection{Simulation Results}

We evaluate the performance of the phase I designs using several
metrics, based on their ability to identify the MTD and the safety in
dose selection and patient allocation. Table 4 summarizes the means
and standard deviations of key performance metrics for the simulation
with 1,000 scenarios. All designs show remarkable similarity with
the largest mean difference across designs only about 0.02. 
This   highlights  
the underlying connection of these designs and echoes our findings
based on the unified decision framework that can generate most designs
as special cases. Figures \ref{table:fixed_scenarios4},
\ref{table:fixed_scenarios5} and \ref{table:fixed_scenarios6} in Appendix D present
the simulation results of the 15 fixed scenarios.  

In general, the five designs tested in the simulation studies exhibit
remarkably similar performances. Specifically, they show comparable
  probabilities (across repeated simulation)  
of allocating patients to the true MTD, and similar risk of
allocating patients to overly toxic doses. The BOIN/CCD and Int-CRM
designs yield slightly higher PCS (probability of correct selection of
MTD) in some cases, such as scenarios 1 and 4 in Figure
\ref{table:fixed_scenarios5}. However, they also   report a  
higher risk in
selecting doses   beyond   the true MTD. For example, in scenarios 2,
3, and 5 in Figure \ref{table:fixed_scenarios5} in Appendix D, the probabilities of
over-dosing selection   under   BOIN, CCD and Int-CRM are higher
compared to the other designs. 

\begin{table}
	 %\centering
 %\def\~{\hphantom{0}}
% \begin{minipage}{130mm}
 \caption{Simulation results of 1,000 random scenarios. Entries are the $\bm{mean}_{sd}$ (across the 1,000 scenarios) proportion of simulated trials for each design and metric.}
	\label{table:random_result}
	\begin{center}
	\begin{tabular}{ccccccc}
        \hline
        Metrics & mTPI & mTPI-2 & BOIN/CCD & Int-CRM & CRM & i3+3 \\ 
  \hline
  Correct Sel. of MTD & $0.60_{(0.15)}$ & $0.62_{(0.14)}$ & $0.62_{(0.14)}$ & $0.62_{(0.14)}$ & $0.62_{(0.14)}$ & $0.62_{(0.14)}$ \\ 
  Sel. over MTD & $0.10_{(0.11)}$ & $0.11_{(0.11)}$ & $0.12_{(0.12)}$ & $0.12_{(0.12)}$ & $0.11_{(0.11)}$ & $0.11_{(0.11)}$ \\ 
 Pat. at MTD & $0.50_{(0.22)}$ & $0.50_{(0.21)}$ & $0.50_{(0.20)}$ & $0.51_{(0.20)}$ & $0.51_{(0.21)}$ & $0.50_{(0.21)}$ \\ 
  Pat. over MTD & $0.11_{(0.10)}$ & $0.11_{(0.10)}$ & $0.12_{(0.11)}$ & $0.12_{(0.11)}$ & $0.12_{(0.10)}$ & $0.11_{(0.10)}$ \\ 
  Tox. & $0.26_{(0.05)}$ & $0.26_{(0.05)}$ & $0.26_{(0.05)}$ & $0.26_{(0.05)}$ & $0.26_{(0.05)}$ & $0.26_{(0.05)}$ \\ 
  None Sel. & $0.04_{(0.07)}$ & $0.04_{(0.07)}$ & $0.04_{(0.07)}$ & $0.04_{(0.07)}$ & $0.04_{(0.07)}$ & $0.04_{(0.07)}$ \\ 
   \hline
	\end{tabular}
	\end{center}
%	\end{minipage}
%	\vskip18pt
\end{table}

%\input{code/table/result_random}

%\clearpage \newpage
\section{Discussion} \label{sec:discussion}

We have   developed   a general decision framework for phase I dose-finding
designs.
  We have shown that   interval-based designs, like mTPI, mTPI-2, BOIN, CCD, and
  the model-based design Int-CRM  
fit into this unified framework.

All   designs use the same 0-1 loss function, and   all
interval designs assume a binomial likelihood function.
  The prior construction for some designs involves the notion of
candidate models.  
Candidate models are
specified assuming different toxic profiles for the doses. Given the
model, the mTPI, mTPI-2 and Int-CRM assume continuous prior
distributions, using beta or normal distributions truncated to the
restricted parameter space implied by the given model. The BOIN and CCD
designs use a different approach
  with a discrete prior on $p_d$, supported at three distinct
values.  
Choosing   those atoms is challenging and  
may be difficult to interpret. However, as shown in many
simulations conducted and published in literature, the BOIN and
CCD designs perform very well in Phase I trials with relatively small
sample size.

Additionally, different loss functions can be considered in the
proposed framework   penalize undesirable actions and outcomes.  
For examples, the loss for mistakenly making an
escalation decision may be larger than for a wrong  
de-escalation. However, such loses usually lead to more complex and
less interpretable decision rules.

It is demonstrated that the designs in this paper perform similarly
with comparable reliability and safety. The i3+3 rule-based design is
not a part of this framework, but also generates similar operating
characteristics, comparable with the other designs.
  The i3+3 design shares a practically important feature with
interval based designs. One can  
pre-tabulate decision tables, which is
a critical feature for   the implementation in actual  
trials. Clinicians can choose 
a desirable design for phase I clinical trial based on their
preference, including the model-based design CRM, the interval
designs, mTPI, mTPI-2, BOIN and CCD, and the rule-based design i3+3.

\bibliographystyle{biom}
\bibliography{bibfile}

\clearpage \newpage
%\newgeometry{top=4cm,bottom=3cm}
\appendix

\section*{Appendix A} 

\noindent \textbf{Safety rules: } Following the mTPI and mTPI-2 designs \citep{ji2010modified, guo2017bayesian}, two safety rules are added, as ethical constraints to avoid excessive toxicity, to all the designs in the simulation study when needed.
\begin{description}
\item[Rule 1: Dose Exclusion.] If the current dose is
  considered excessively toxic, i.e., $p\{ p_d>p_T\mid Data \}>\xi$,
  where the threshold $\xi$ is close to 1, say 0.95, then the current and all
  higher doses are excluded and never used again for the remainder of
  the trial.
\item[Rule 2: Early Stopping.] If the current
  dose is the lowest dose (first dose) and is considered
  excessively toxic according to Rule 1,   then stop the   trial
  for safety.    
\end{description}
In safety Rules 1 and 2, $Prob\{ p_d>p_T\mid Data \}$ is a function of the cumulative beta distribution $\Be(\alpha_0+y_d,\beta_0+n_d-y_d)$, and $\alpha_0=\beta_0=1$ is used by default.

\section*{Appendix B} 
\noindent \textbf{Intervals in Int-CRM design: }
The intervals in the Int-CRM design are,
$$ I_k = \{\theta \in [A_1,A_{ T+1}]: |F(k,\theta) - p_T| < |F(d,\theta) - p_T|, \forall d \neq k \}, \;\;\; k=1, \dots , T, $$
which have the form $I_1 = [\psi_1 = A_1, \psi_2)$, $I_k = [\psi_k,\psi_{k+1}),
\; k=2,...,( T-1)$, $I_{ T} = [\psi_ T,\psi_{ T+1} = A_{ T+1}]$. 
The boundary of the intervals $\psi_k$ satisfies the equation
$$ F(k-1,\psi_k) + F(k, \psi_k) = 2p_T, \;\;\; k=2,..., T.$$
Under the model of Int-CRM,
$$ q_{k-1}^{exp(\phi_k)} + q_{k}^{exp(\phi_k)} = 2 p_T.$$
Therefore, given the skeleton $(q_1,...,q_T)$ and $p_T$, we can obtain the numerical solution of the interval boundary by searching over a sequence of $\phi_k \in [A_1,A_{ T+1}]$.

\clearpage \newpage 

\section*{Appendix C}
\subsection*{Proof of Proposition 1}
\begin{proof}

By definition,

% \begin{eqnarray*}
\begin{eqnarray*}
\RR_d &=& \arg \min_{a \in \AA} \int \ell(a,p_d) p(p_d \mid \bm{y}) dp_d 
= \arg \min_{a \in \AA} \int  \delta(p_d \notin I_a)  p(p_d \mid \bm{y}) dp_d \\
&=&\arg \min_{a \in \AA} \int \{ 1-\delta(p_d \in I_a)  \} p(p_d \mid \bm{y}) dp_d 
= \arg \max_{a \in \AA} \int \delta(p_d \in I_a) p(p_d \mid \bm{y}) dp_d\\
&=& \arg \max_{a \in \AA} Pr(p_d \in I_a \mid \bm{y}),
 \nonumber
\end{eqnarray*}
which proves the first equation in Proposition 1. In addition, we have
\begin{eqnarray} \label{eq:pfthe1_1}
\RR_d & =& \arg \max_{a \in \AA} \int \sum_{k=1}^3 \delta(p_d \in I_a) p(\bm{y} \mid p_d)\pi(p_d \mid m=k)\pi(m=k) dp_d 
\end{eqnarray}
Since $$\AA = \{E,S,D\}, I_1 = I_E, I_2 = I_S, I_3 = I_D,$$ 
equation \ref{eq:pfthe1_1} becomes
\begin{eqnarray*}
\RR_d & =& \arg \max_{a \in \AA} \int \sum_{k=1}^{3} \delta(p_d \in I_a) p(\bm{y} \mid p_d) g(p_d) \delta(p_d \in I_k) \pi(m=k) dp_d \\
& =& \arg \max_{k \in \{1,2,3\}} \int p(\bm{y} \mid p_d) g(p_d) \delta(p_d \in I_k) \pi(m=k) dp_d \\
& =& \arg \max_{k \in \{1,2,3\}} \int  p(\bm{y} \mid p_d)\pi(p_d \mid m=k)\pi(m=k) dp_d \\
& =& \arg \max_{k \in \{1,2,3\}} \int  p(\bm{y} \mid p_d)\pi(p_d \mid m=k) dp_d \\
& = &\arg \max_{k \in \{1,2,3\}} Pr(m=k \mid \bm{y})
\end{eqnarray*}

The penultimate equation is true since $\pi(m=k) = \frac{1}{3}, \; k=1,2,3.$

\end{proof}

\subsection*{Proof of Corollary	\ref{cor:mtpi}}
\begin{proof}
  Recall the action space $\AA = \{E,S,D\}$. Based on Equation
  (\ref{eq:the1}) in Proposition 1, the Bayes' rule is
  \begin{eqnarray*}
	\RR_d  &=& \arg \max_{k \in \{1,2,3\}} Pr(m=k \mid y)  \\
	 &=& \arg \max_{k \in \{1,2,3\}} \int  p(y \mid p_d)\pi(p_d \mid m=k) dp_d \\
	 &=& \arg \max_{a \in \AA} \int p_d^{y_d} (1-p_d)^{n_d-y_d} \cdot C_k beta(p_d;1,1)\delta(p_d \in I_a) dp_d \\
	 &=& \arg \max_{a \in \AA} \int p_d^{y_d} (1-p_d)^{n_d-y_d} \frac{beta(p_d;1,1) \delta(p_d \in I_a)}{\int_{I_a} beta(p;1,1) dp} dp_d \\
	 &=& \arg \max_{a \in \AA} \int p_d^{y_d} (1-p_d)^{n_d-y_d} \frac{\delta(p_d \in I_a)}{||I_a||} dp_d 
	 = \arg \max_{a \in \AA} \UPM(I_a) 
	 = R_{mTPI}
        \nonumber
  \end{eqnarray*}        
\end{proof}

\subsection*{Proof of Corollary	\ref{cor:mtpi2}}

\begin{proof}
	Based on Equation (\ref{eq:the1}), the Bayes' rule is equal to
	\begin{eqnarray*}
	\RR_d & =& \arg \max_{m \in \AA_m} Pr(m=k \mid y)  \\
	 &=& \arg \max_{m \in \{1,...,K\}} \int  p(y \mid p_d)\pi(p_d \mid m=k) dp_d \\
	& =& \arg \max_{m \in \{1,...,K\}} \int p_d^{y_d} (1-p_d)^{n_d-y_d} \cdot C_k beta(p_d;1,1)\delta(p_d \in I_m) dp_d \\
	& =& \arg \max_{m \in \{1,...,K\}} \int p_d^{y_d} (1-p_d)^{n_d-y_d} \frac{beta(p_d;1,1) \delta(p_d \in I_m)}{\int_{I_m} beta(p;1,1) dp} dp_d \\
	& =& \arg \max_{m \in \{1,...,K\}} \int p_d^{y_d} (1-p_d)^{n_d-y_d} \frac{\delta(p_d \in I_m)}{||I_m||} dp_d \\
	 &=& \arg \max_{m \in \AA_m} \UPM(I_m).
	\end{eqnarray*}
	
\end{proof}

\subsection*{Proof of Corollary \ref{cor:ccd}}
\begin{proof}
	Based on Equation (\ref{eq:the1}) and the proof of Theorem \ref{theorem2}, the Bayes' rule is
	\begin{eqnarray*}
	\RR_d & =& \arg \max_{a \in \{E,S,D\}}  {\phi'_a}^{y_d} (1-{\phi'_a})^{n_d-y_d},
	% \end{eqnarray*} % and
	% \begin{eqnarray*} % \RR_d =
	 \left\{\begin{array}{ll} E, & \hat{p}_d \leq \lambda_1, \\ S, & \lambda_1
< \hat{p}_d < \lambda_2, \\ D, & \hat{p}_d \geq \lambda_2 ,
	\end{array}\right.
	\end{eqnarray*} with $ \lambda_1 = \xi(\phi_E) = \xi(\xi^{-1}
(p_T-\epsilon_1)) = p_T - \epsilon_1$ and $\lambda_2 = \xi(\phi_D) =
\xi(\xi^{-1} (p_T + \epsilon_2)) = p_T + \epsilon_2$.  Therefore,
	\begin{eqnarray*} \RR_d &=&
	 \left\{\begin{array}{ll} E & \hat{p}_d \leq p_T - \epsilon_1 \\ S & p_T -
\epsilon_1 < \hat{p}_d < p_T + \epsilon_2 \\ D & \hat{p}_d \leq p_T +
\epsilon_2
	\end{array}\right. \hspace{1cm} = \RCCD.
	\end{eqnarray*}
	
\end{proof}

\subsection*{Proof of Theorem \ref{theorem2}}
\begin{proof} 
	According to Proposition 1, under the proposed decision framework, the Bayes' rule is
	\begin{eqnarray}
	\RR_d & =& \arg \max_{k \in \{1,2,3\}} Pr(m=k \mid y)  
	 = \arg \max_{k \in \{1,2,3\}} \int  p(y \mid p_d)\pi(p_d \mid m=k) dp_d \nonumber \\
	& = &\arg \max_{k \in \{1,2,3\}} \int p_d^{y_d} (1-p_d)^{n_d-y_d} \delta{(p_d = \phi_k)} dp_d \nonumber \\
	& = &\arg \max_{k \in \{1,2,3\}}  {\phi_k}^{y_d} (1-{\phi_k})^{n_d-y_d} 
	 = \mbox{arg} \max_{a \in \{E,S,D\}} {\phi_a}^{y_d} {(1-\phi_a)}^{n_d-y_d}.
	\label{eq:BOIN}
	\end{eqnarray}
	Let $ g(p) = p^{y_d} (1-p)^{n_d-y_d},$ then equation \ref{eq:BOIN} becomes
	$$\RR_d = \mbox{arg} \max_{a \in \{E,S,D\}} g(\phi_a).$$
	When $y_d = 0$,
	$g(p) = (1-p)^{n_d},$ and $g(p)$ is monotonically decreasing with $p$. When $y_d >0$,
	$$ \frac{dg(p)}{dp} = (y_d-n_dp) \left \{y_dp^{(y_d-1)}(1-p)^{n_d - y_d -1} \right \}.$$
	Note that the term
	$\left \{y_dp^{y_d-1}(1-p)^{n_d - y_d -1} \right \} >0$. Then, 
	$\frac{dg(p)}{dp}> 0$ if $p \in (0,\frac{y_d}{n_d})$; and $\frac{dg(p)}{dp}< 0$ if $p \in (\frac{y_d}{n_d},1)$. Therefore,
	the function $g(p)$ first increases and then decreases with $p$. It only has one mode which is a maximum. In summary, $g(p)$ either monotonically decreases or first increases and decreases with $p$.
	
	$1)$ When $\hat{p}_d \leq \lambda_1$,
	$$\frac{y_d}{n_d} \leq \lambda_1 = \frac{log\left( \frac{1-\phi_E}{1-\phi_S}\right) }{log\left( \frac{\phi_S(1-\phi_E)}{\phi_E(1-\phi_S)}\right) } 
	= \frac{log(1-\phi_E) - log(1-\phi_S)}{log(\frac{\phi_S}{1-\phi_S}) - log(\frac{\phi_E}{1-\phi_E})}.$$
	
	And because $0<\phi_E<\phi_S<\phi_D<1$, we have $log(\frac{\phi_E}{1-\phi_E}) < log(\frac{\phi_S}{1-\phi_S}) < log(\frac{\phi_D}{1-\phi_D}).$ Therefore, $log(\frac{\phi_S}{1-\phi_S}) - log(\frac{\phi_E}{1-\phi_E}) > 0.$ Hence,
	$$y_d \left \{log(\frac{\phi_S}{1-\phi_S}) - log(\frac{\phi_E}{1-\phi_E}) \right \} \leq n_d \left \{ log(1-\phi_E) - log(1-\phi_S) \right \},$$
	and by taking exponentiation on both sides, we have
	$$ (\frac{\phi_S}{1-\phi_S})^{y_d}(1-\phi_S)^{n_d} \leq (\frac{\phi_E}{1-\phi_E})^{y_d}(1-\phi_E)^{n_d}, $$
	which leads to
	$$ \phi_S^{y_d}(1-\phi_S)^{n_d-y_d} \leq \phi_E^{y_d}(1-\phi_E)^{n_d-y_d}, $$
	i.e.,
	$g(\phi_S) \leq g(\phi_E).$
	Since $g(p)$ either monotonically decreases or first increases and then decreases with $p$, and because $\phi_E<\phi_S<\phi_D$ and $g(\phi_E) \geq g(\phi_S)$, we have
	$g(\phi_E) \geq g(\phi_S) > g(\phi_D).$
	Therefore, when $\hat{p}_d \leq \lambda_1$, 
	$\RR_d = \mbox{arg} \max_{a \in \{E,S,D\}} g(\phi_a) = E.$
	
	$2)$ When $\hat{p}_d \geq \lambda_2$,
	$$\frac{y_d}{n_d} \geq \lambda_2 = \frac{log\left( \frac{1-\phi_D}{1-\phi_S}\right) }{log\left( \frac{\phi_S(1-\phi_D)}{\phi_D(1-\phi_S)}\right) } 
	= \frac{log(1-\phi_D) - log(1-\phi_S)}{log(\frac{\phi_S}{1-\phi_S}) - log(\frac{\phi_D}{1-\phi_D})}.$$
	Since $log(\frac{\phi_S}{1-\phi_S}) - log(\frac{\phi_D}{1-\phi_D}) <0$, 
	$$y_d \left \{log(\frac{\phi_S}{1-\phi_S}) - log(\frac{\phi_D}{1-\phi_D}) \right \} \leq n_d\left \{log(1-\phi_D) - log(1-\phi_S)\right \},$$
	and
	$$ (\frac{\phi_S}{1-\phi_S})^{y_d}(1-\phi_S)^{n_d} \leq (\frac{\phi_D}{1-\phi_D})^{y_d}(1-\phi_D)^{n_d}, $$
	which leads to
	$$ \phi_S^{y_d}(1-\phi_S)^{n_d-y_d} \leq \phi_D^{y_d}(1-\phi_D)^{n_d-y_d}, $$
	i.e.,
	$g(\phi_S) \leq g(\phi_D).$
	Again, either $g(p)$ monotonically decreases or first increases then decreases with $p$. In either cases, we have
	$g(\phi_E) < g(\phi_S) \leq g(\phi_D).$
	Therefore, when $\hat{p}_d \geq \lambda_2$, 
	$\RR_d = \mbox{arg} \max_{a \in \{E,S,D\}} g(\phi_a) = D.$
	
	$3)$ When $\lambda_1 < \hat{p}_d < \lambda_2$,
	we have 
	$$ \phi_S^{y_d}(1-\phi_S)^{n_d-y_d} > \phi_E^{y_d}(1-\phi_E)^{n_d-y_d}, $$
	and
	$$ \phi_S^{y_d}(1-\phi_S)^{n_d-y_d} > \phi_D^{y_d}(1-\phi_D)^{n_d-y_d}, $$
	i.e.,
	$g(\phi_S) > g(\phi_E)$ and 
	$g(\phi_S) > g(\phi_D).$
	Therefore, when $\lambda_1 < \hat{p}_d < \lambda_2$, 
	$\RR_d = \mbox{arg} \max_{a \in \{E,S,D\}} g(\phi_a) = S.$
	
\end{proof}

\clearpage \newpage
\section*{Appendix D}

Operating characteristics for the 15 fixed scenarios in the simulation study. The yellow bar highlights the true MTD.
{\em Prob of Select at/over MTD} refers to the probability (over
   repeat simulation) of selecting the true MTD and a dose above the
   true MTD, respectively.
   {\em Prob of Pat. at/over MTD} refers to the relative frequency of
   patients assigned at or above the true MTD.
    {\em Prob of Toxicity} refers to the frequency of patients experienced DLT in all simulated trials.
   {\em Prob of no selection} refers to the probability of failing to
   recommend any dose.

\begin{figure}
	\centering
	\caption{Simulation results of the 5 fixed scenarios with 4 doses levels. }
	\label{table:fixed_scenarios4}
	\centerline{\includegraphics[width=5.6in]{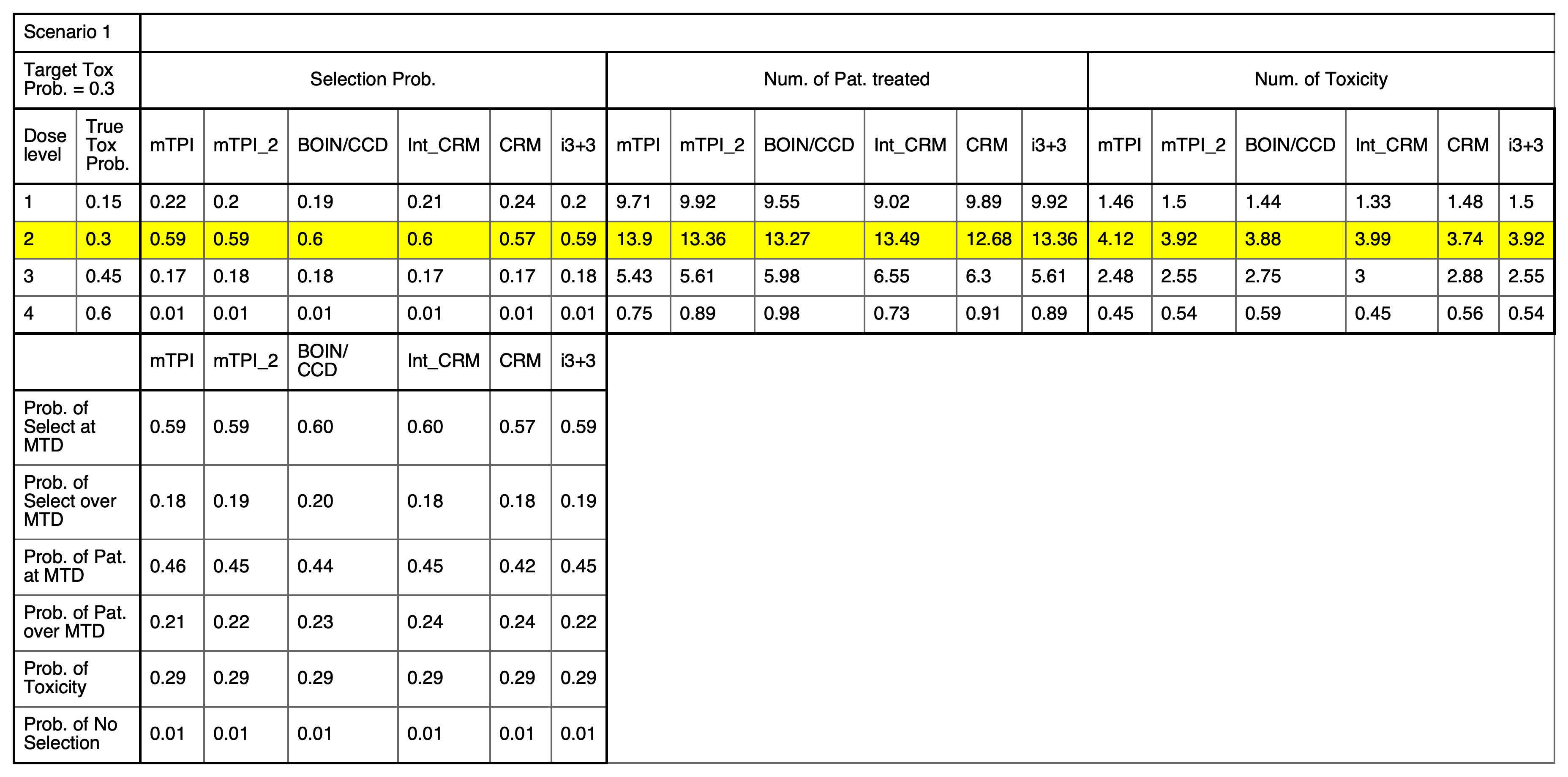}}
	
	\centerline{\includegraphics[width=5.6in]{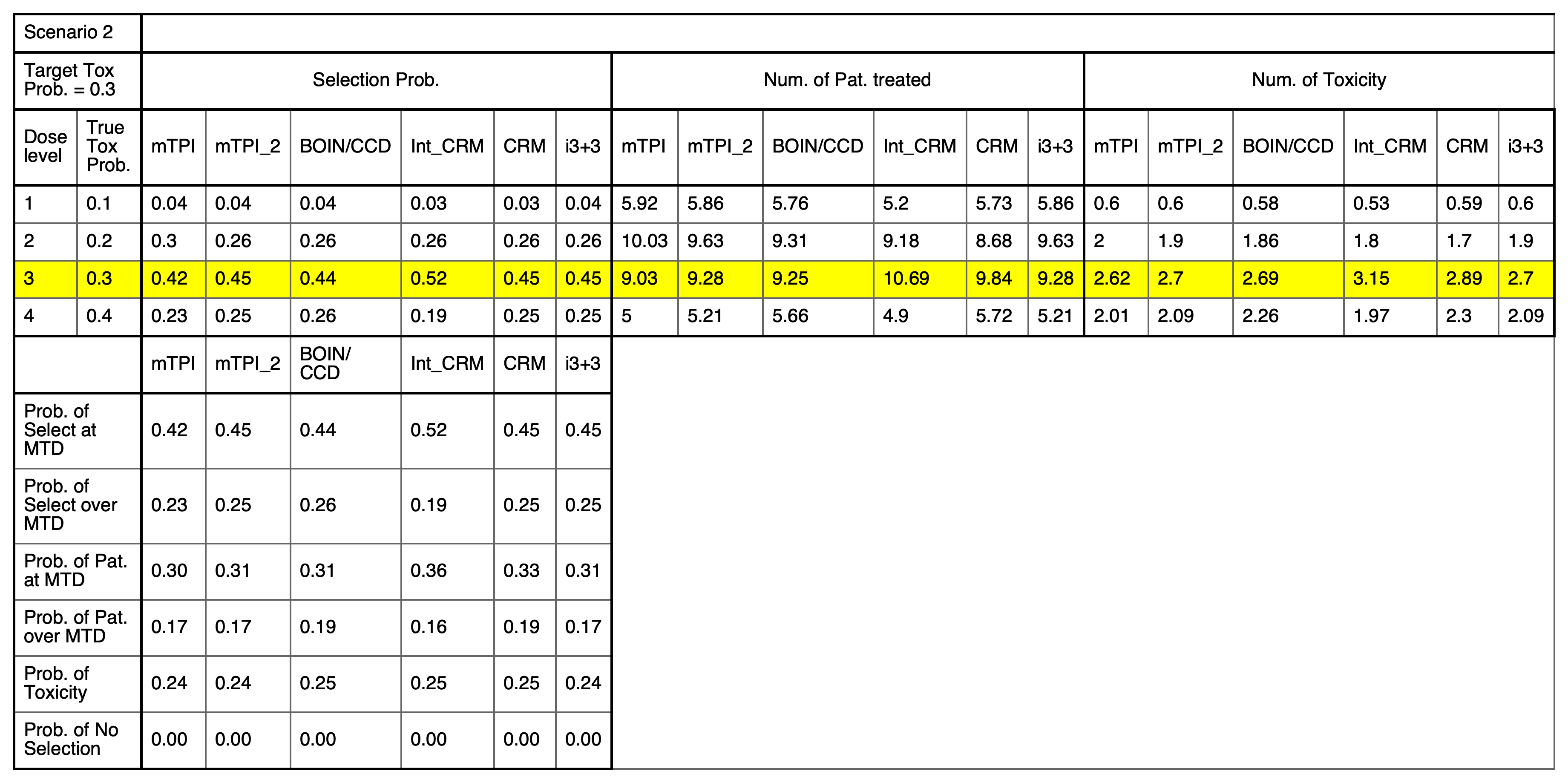}}
	
	\centerline{\includegraphics[width=5.6in]{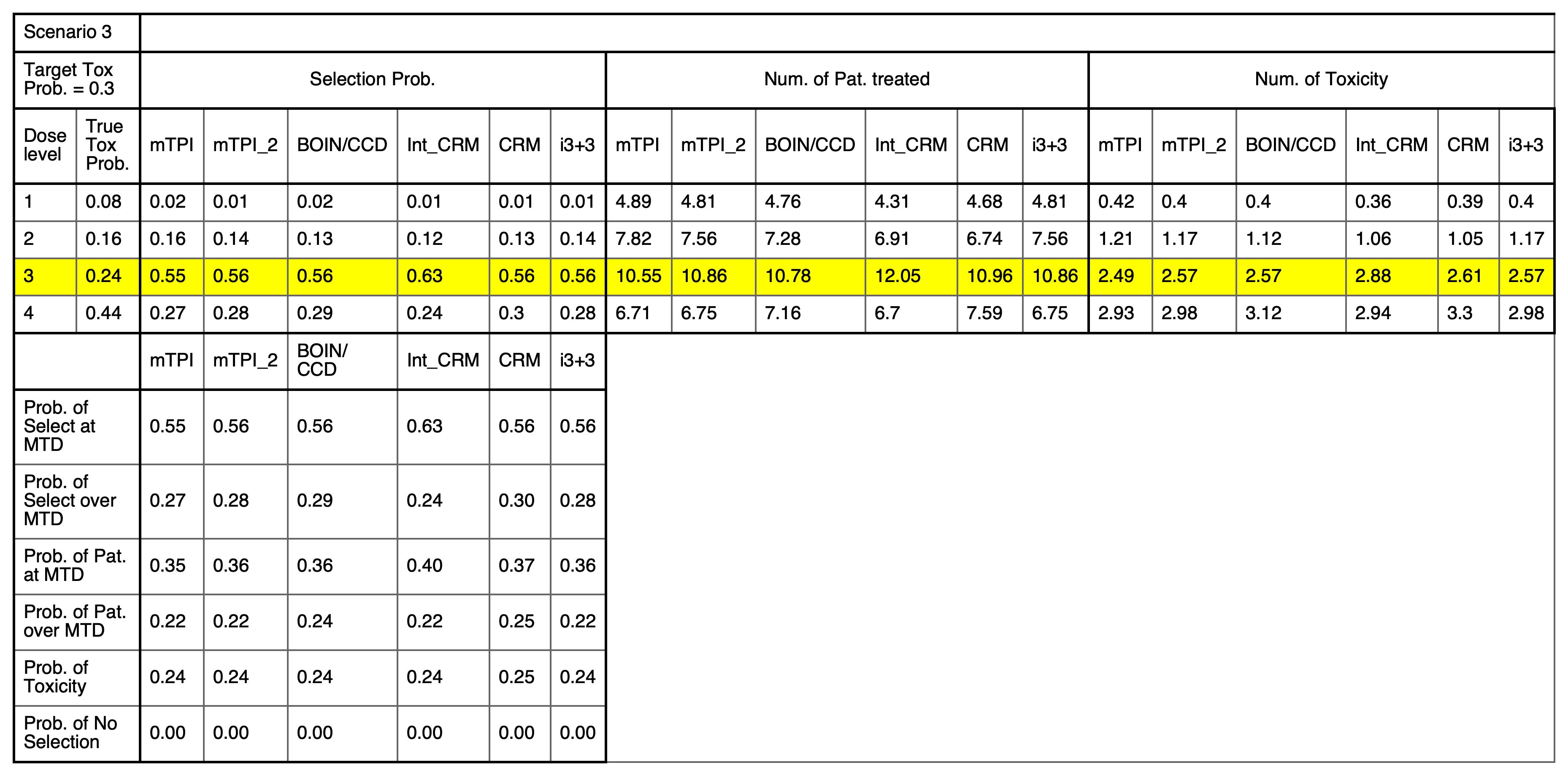}}
\end{figure} 

\begin{figure}
	\centering
	\centerline{\includegraphics[width=5.6in]{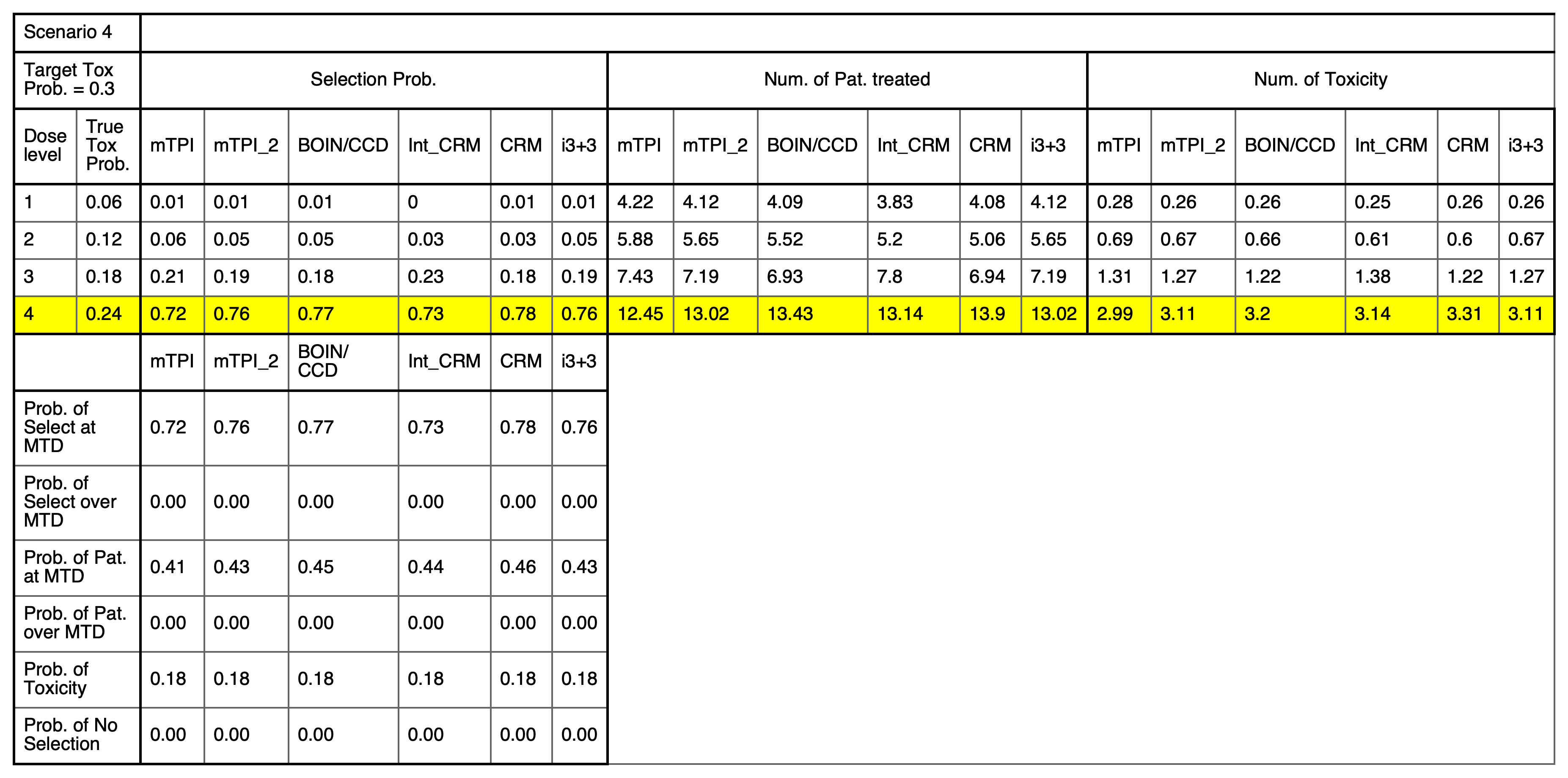}}
	\centerline{\includegraphics[width=5.6in]{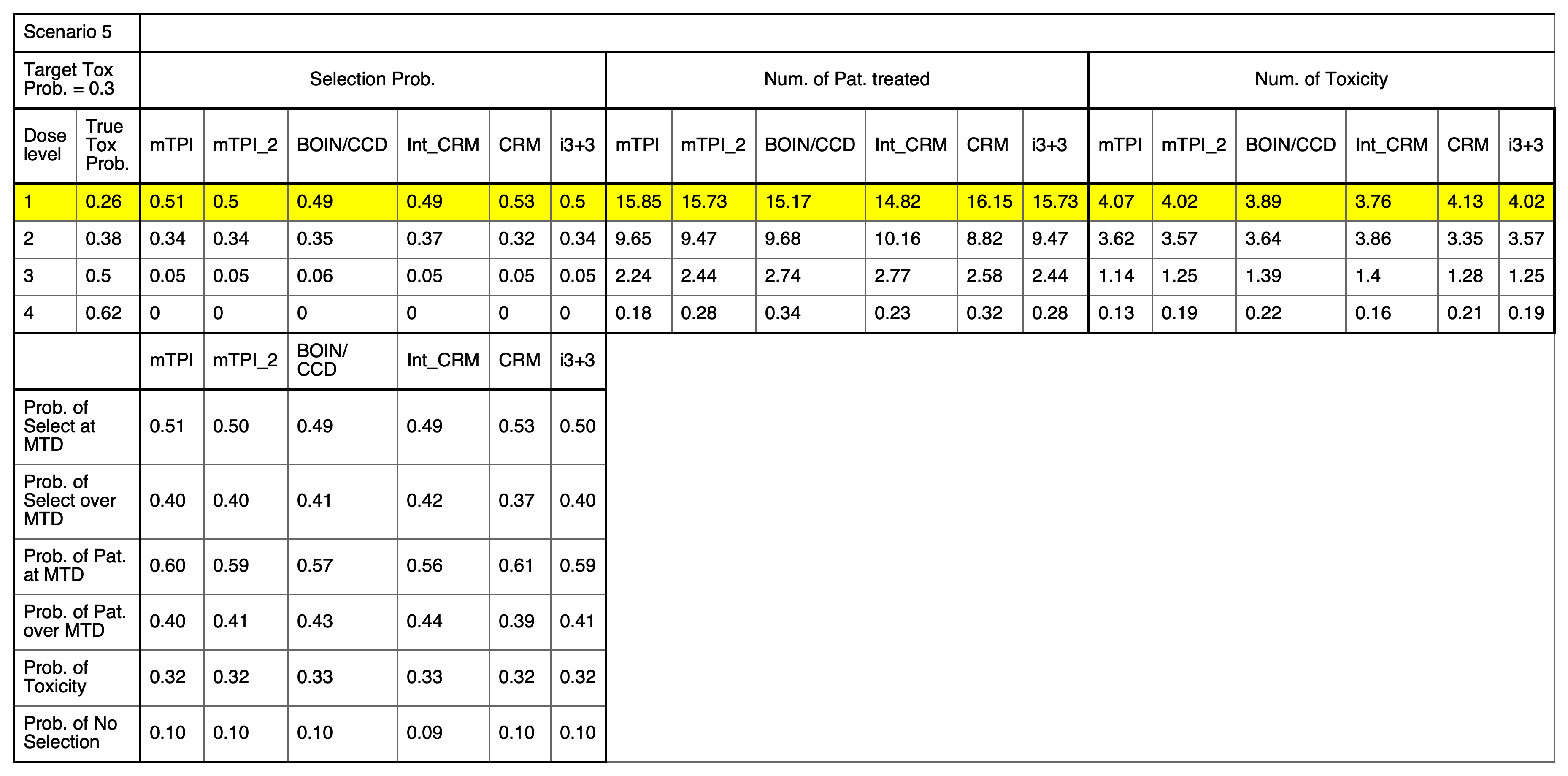}}
\end{figure} 

\begin{figure}
	\centering
	\caption{Simulation results of the 5 fixed scenarios with 4 doses levels. }
	\label{table:fixed_scenarios5}
	\centerline{\includegraphics[width=5.6in]{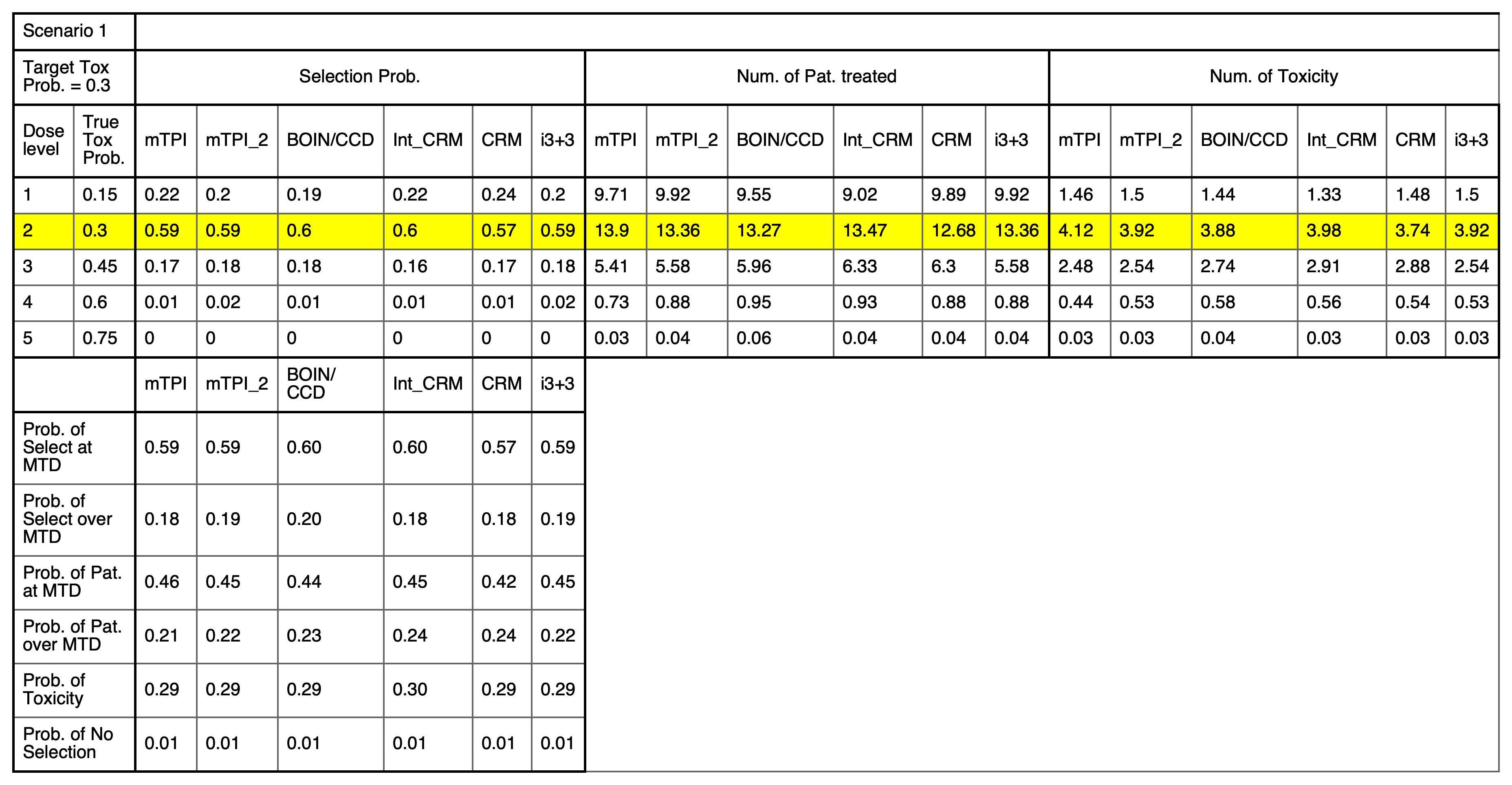}}
	
	\centerline{\includegraphics[width=5.6in]{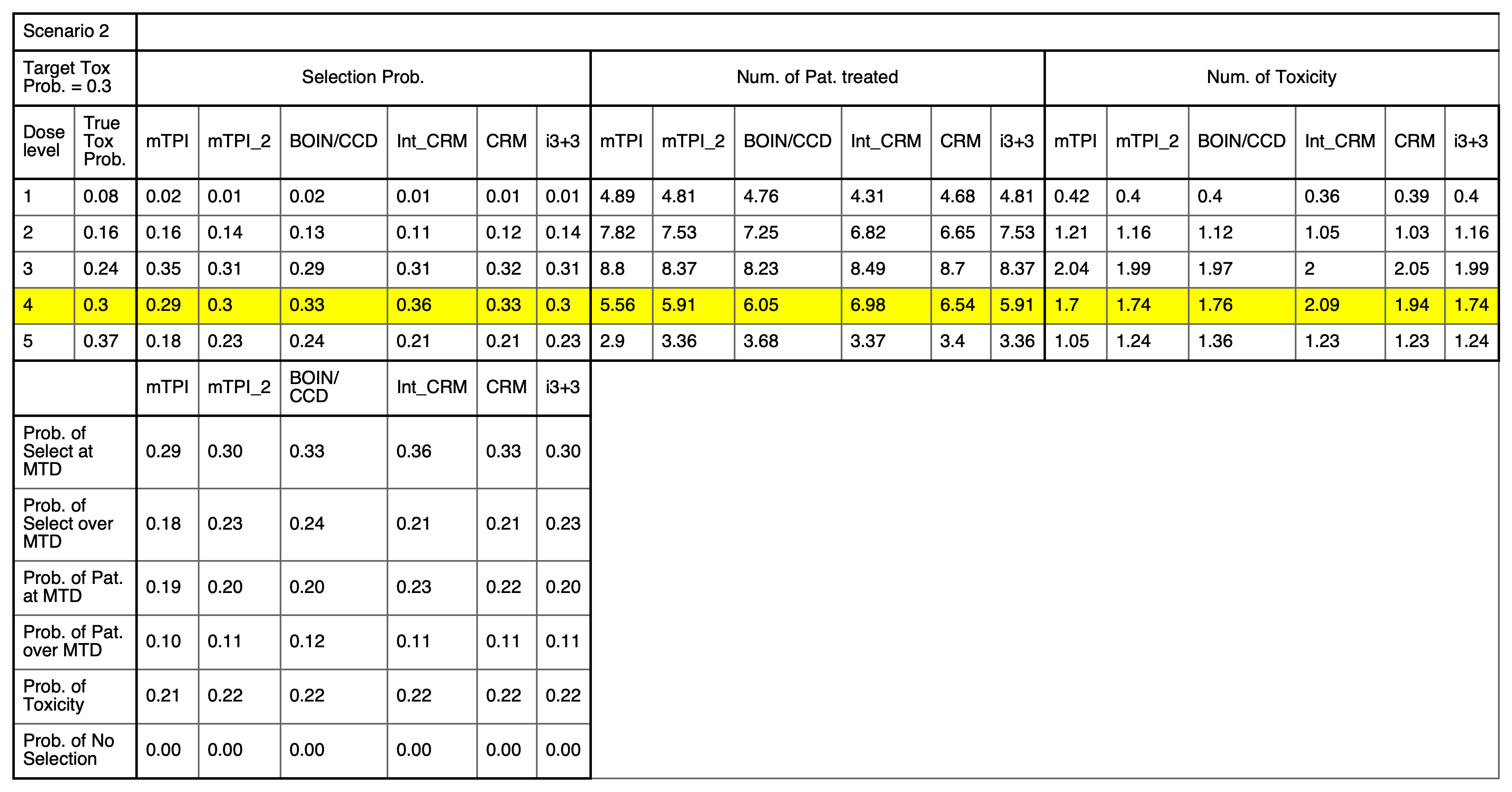}}
	
	\centerline{\includegraphics[width=5.6in]{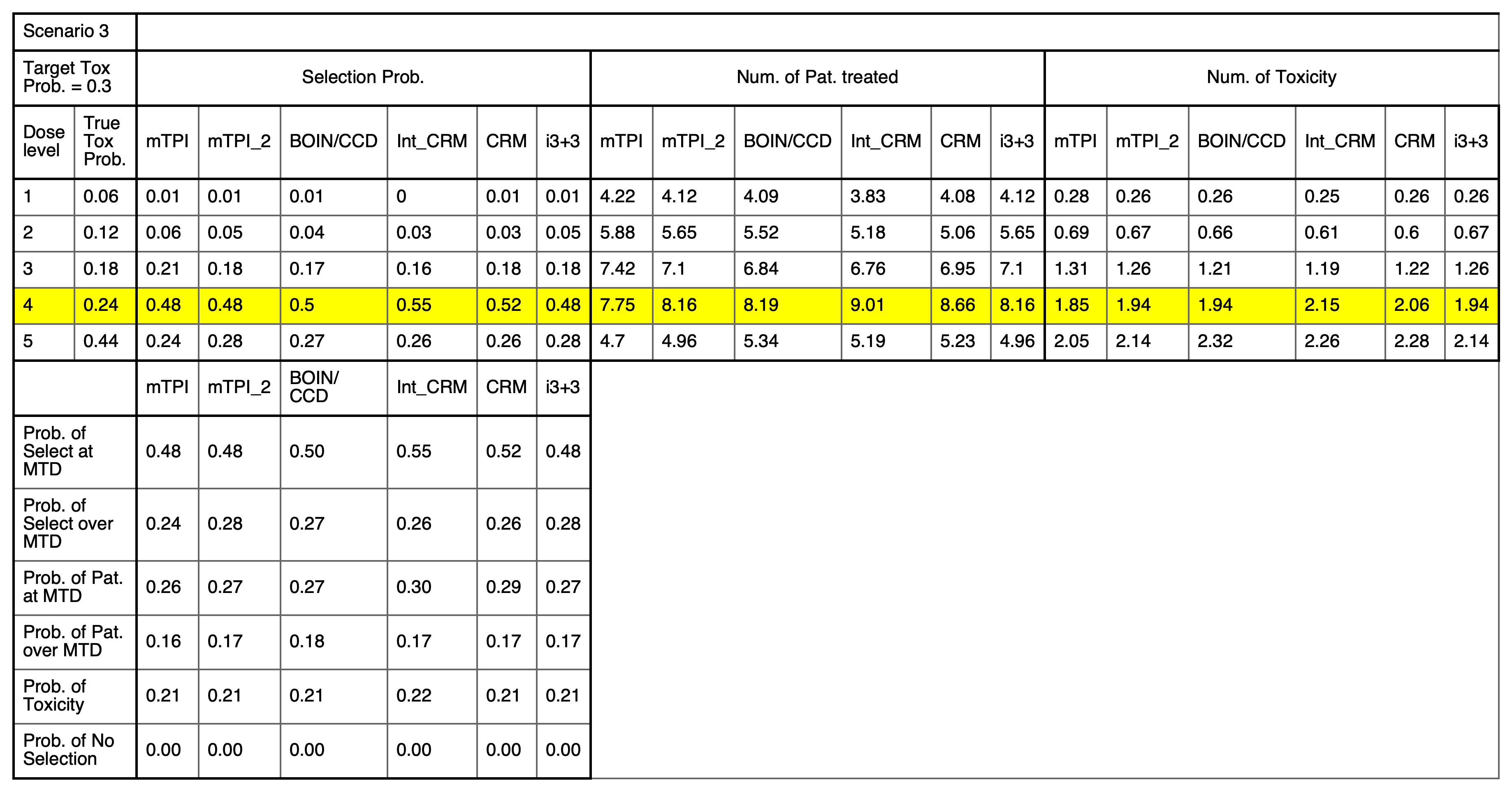}}
\end{figure} 

\begin{figure}
	\centering
	\centerline{\includegraphics[width=5.6in]{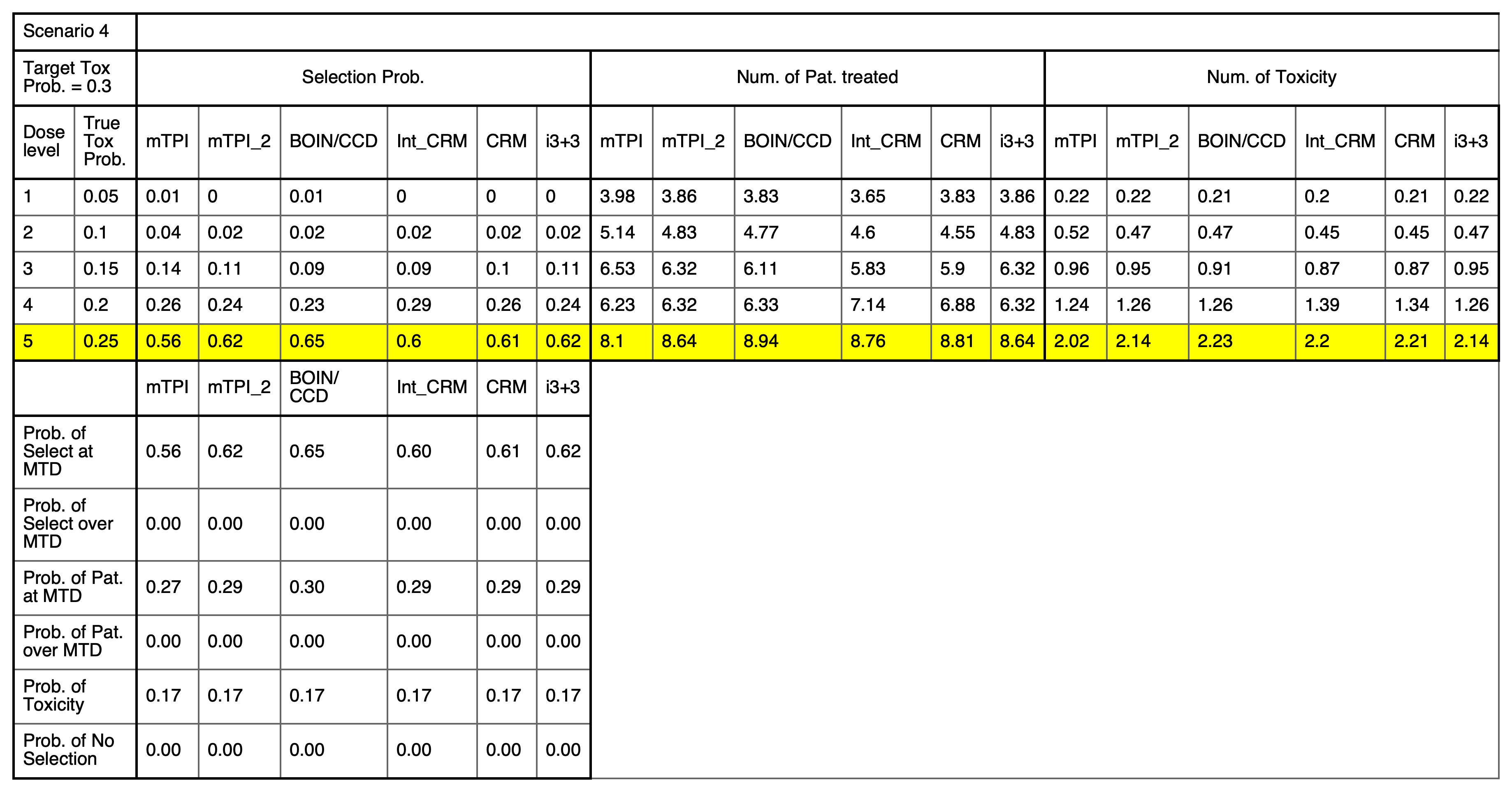}}
	\centerline{\includegraphics[width=5.6in]{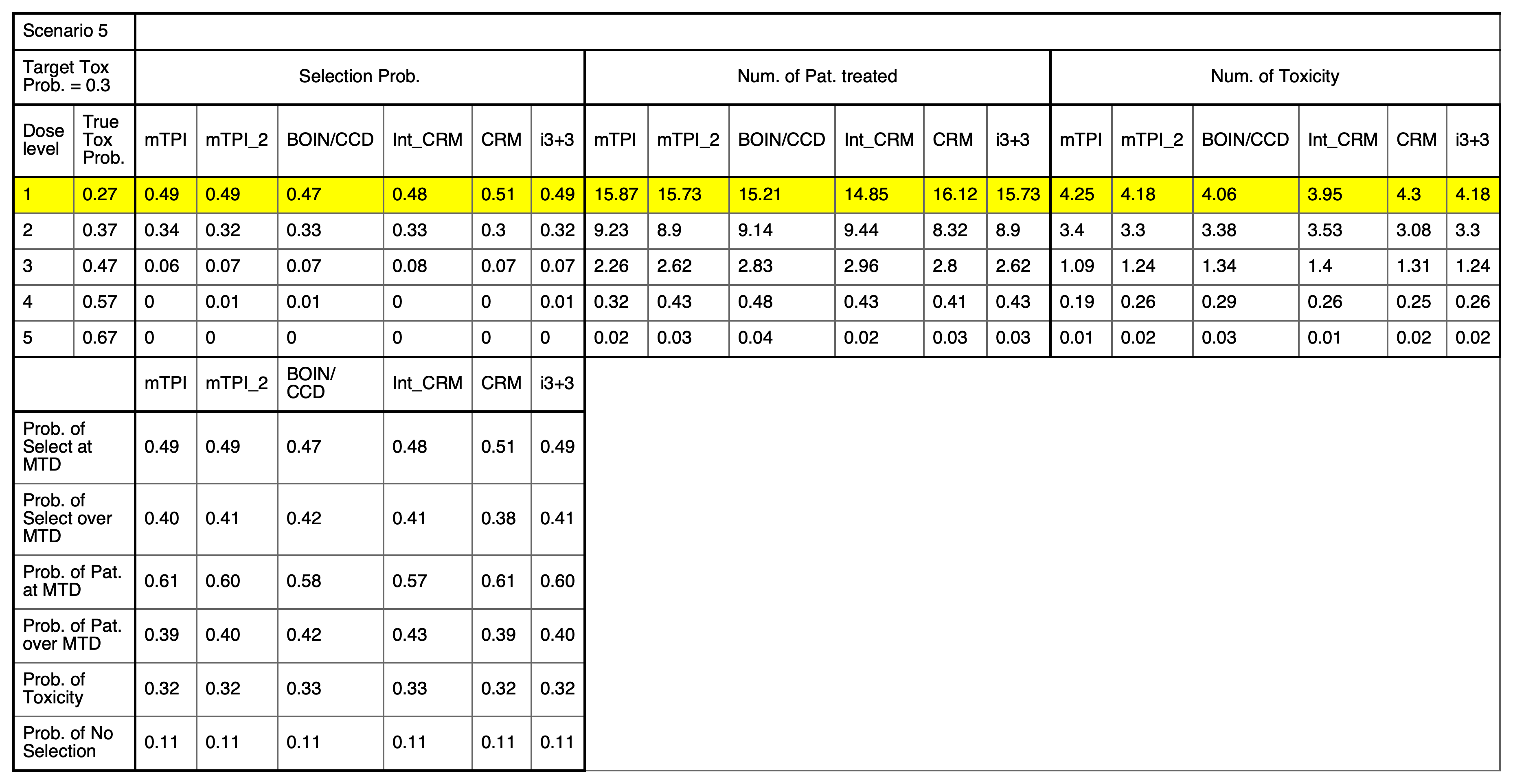}}
\end{figure} 

\begin{figure}
	\centering
	\caption{Simulation results of the 5 fixed scenarios with 4 doses levels. }
	\label{table:fixed_scenarios6}
	\centerline{\includegraphics[width=5.6in]{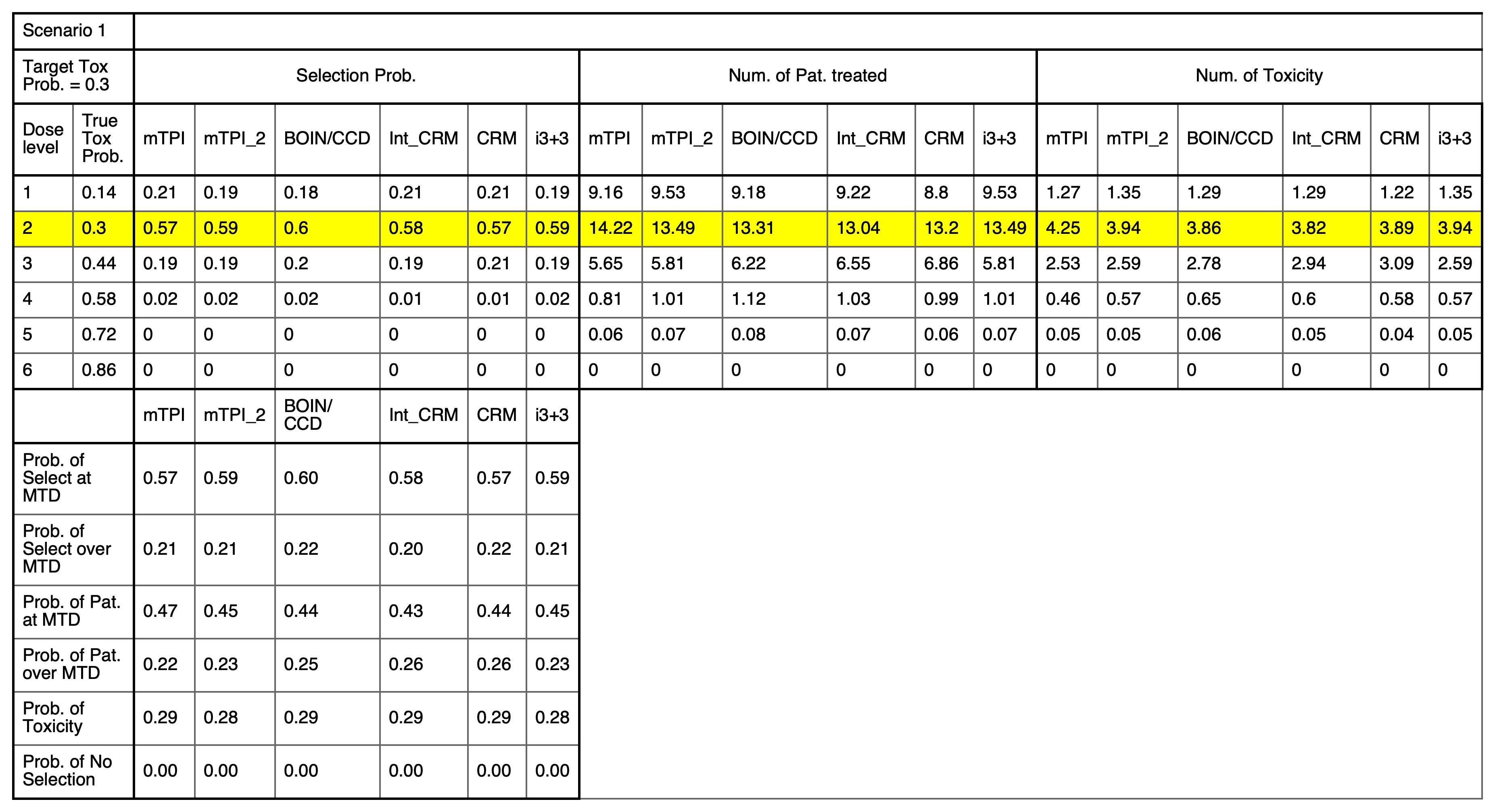}}
	
	\centerline{\includegraphics[width=5.6in]{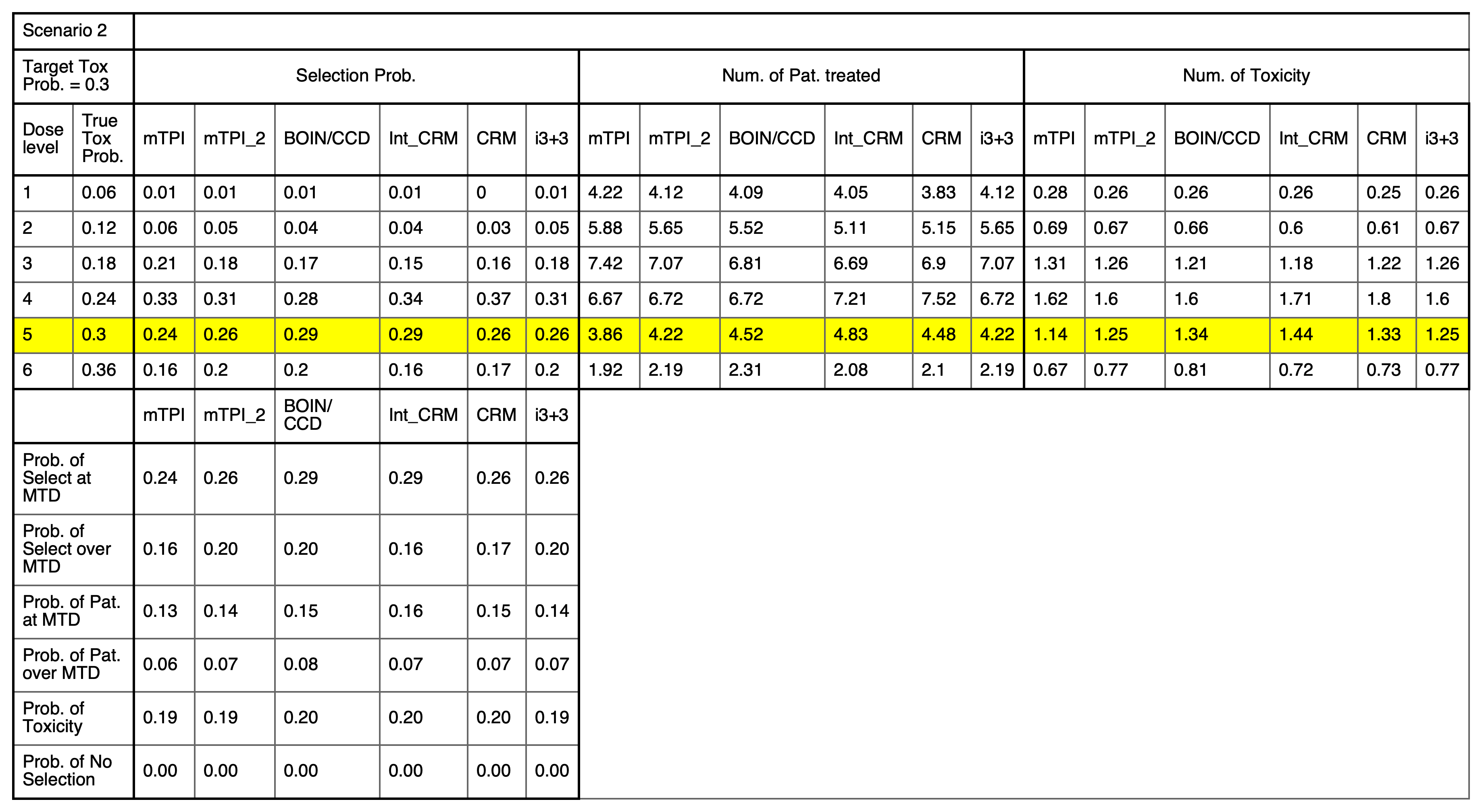}}
	
	\centerline{\includegraphics[width=5.6in]{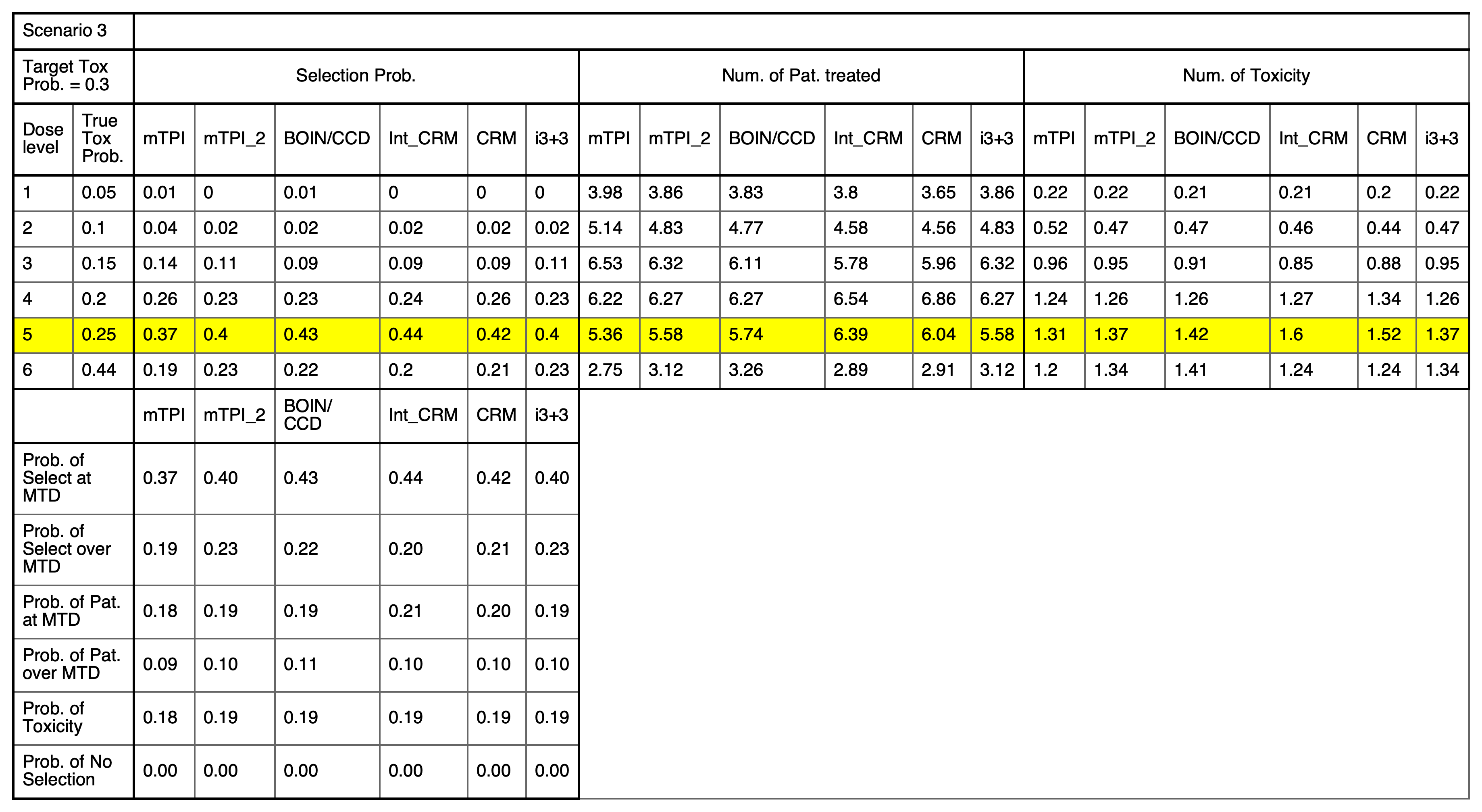}}
\end{figure} 

\begin{figure}
	\centering
	\centerline{\includegraphics[width=5.6in]{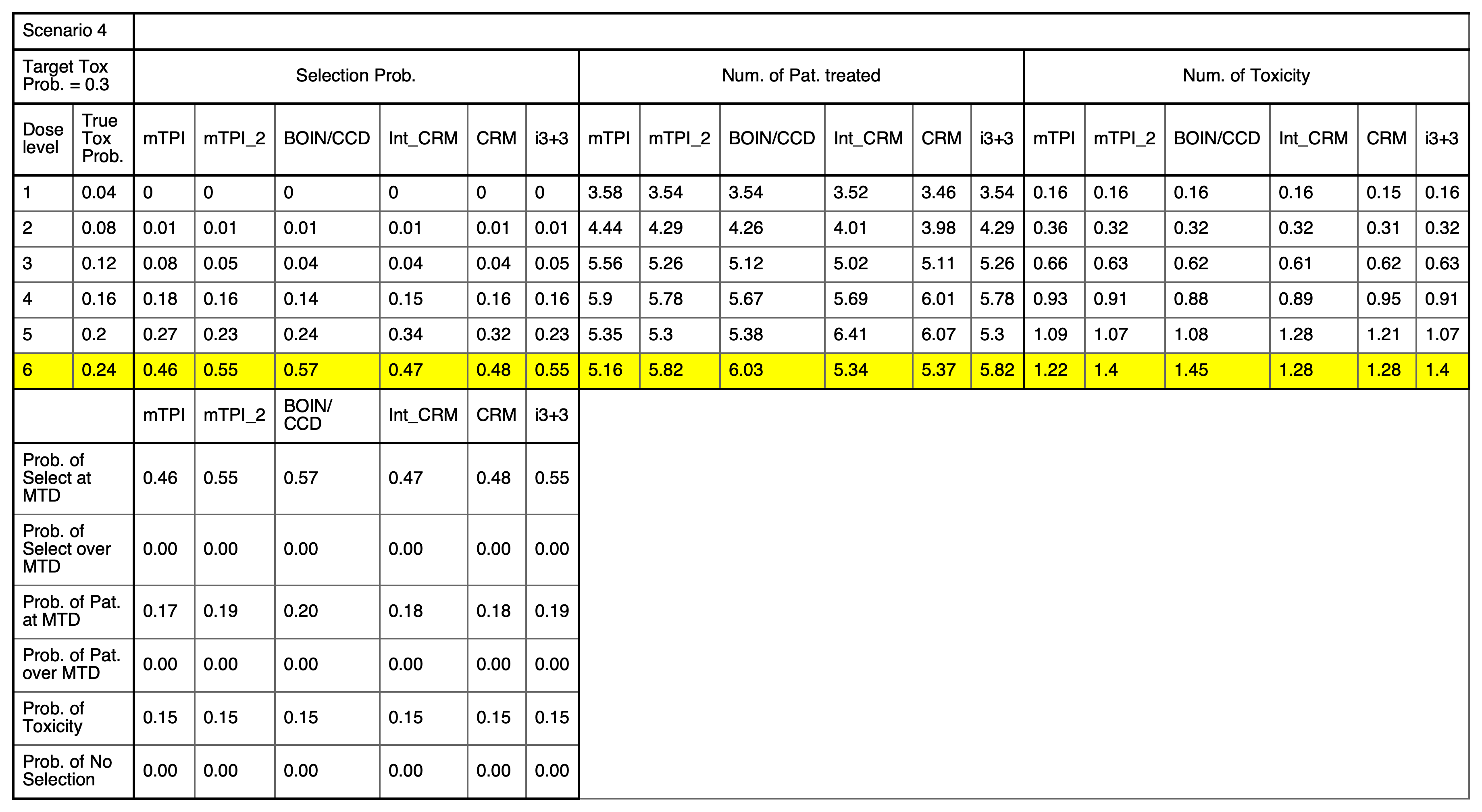}}
	\centerline{\includegraphics[width=5.6in]{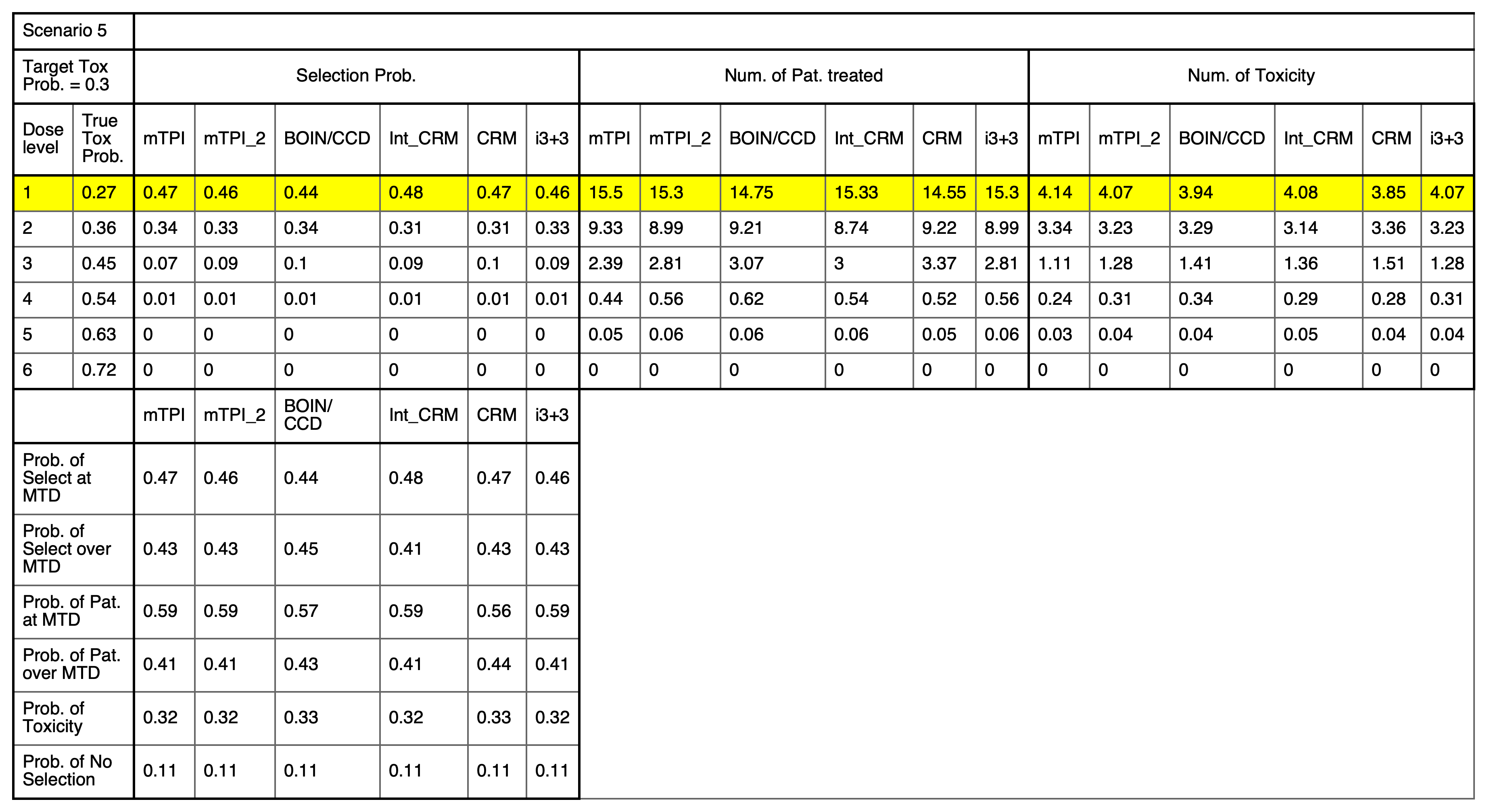}}
\end{figure}

\end{document}